\documentclass[
 aip,
 cha,
 amsmath,amssymb,
 reprint,
 superscriptaddress,
 nolongbibliography
]{revtex4-1}
\usepackage[T1]{fontenc}
\usepackage[utf8]{inputenc}
\usepackage{graphicx,xcolor}
\usepackage{dcolumn}
\usepackage{bm}
\usepackage{amsfonts}
\usepackage{etoolbox}
\begin{document}

\title{Birth and breakdown of breathing and rotobreathing cyclops states in Kuramoto networks with higher-mode coupling}

\author{Matvei M. Khamkov}
\author{Maxim I. Bolotov}
\author{Lev A. Smirnov}
\affiliation{Department of Control Theory and Research and Educational Center for Mathematics of Future Technologies, Lobachevsky State University of Nizhny Novgorod, 23 Gagarin Avenue, Nizhny Novgorod 603022, Russia}

\author{Igor Belykh}
\email[Corresponding author: ]{ibelykh@gsu.edu}
\affiliation{Department of Mathematics and Statistics and Neuroscience Institute, Georgia State University, P.O. Box 4110, Atlanta, Georgia 30302-410, USA}
    
\begin{abstract}
Cluster synchronization states often serve as organizing centers for collective dynamics, but their destabilization can open unexpected routes to both coherent and incoherent behavior. We study such transitions for cyclops states in globally coupled networks of identical Kuramoto-Sakaguchi rotators with inertia and two-harmonic coupling. Stationary cyclops states consist of two coherent clusters and a solitary oscillator that maintains fixed phase differences with the clusters. Using Floquet analysis and numerical continuation of periodic 
orbits, we trace bifurcation routes for the birth and breakdown of breathing and rotobreathing cyclops states with nonstationary intercluster phase differences. Breathing cyclops states, born from their stationary counterparts, correspond to bounded oscillations of the intercluster phases. They lose stability via period-doubling bifurcations, which produce phase-split cyclops states whose intercluster motion repeats only after two cycles of the parent breather, or via cluster-destruction bifurcations. 
Rotobreathing cyclops states, in which the intercluster phase differences undergo full rotations, are not merely the large-amplitude continuation of breathing cyclops states; instead, they form a separate family of nonstationary cyclops dynamics born through global bifurcations involving heteroclinic-contour-like structures of saddle cluster states.
We further show that these states have wide, often global, basins of attraction, persist in large odd-sized networks, and contrast sharply with even-sized networks, where stationary multi-cluster states dominate. These results identify higher-harmonic coupling and solitary-oscillator-mediated rotations as generic mechanisms for organizing complex cluster motion in phase oscillator networks.
\end{abstract}

\keywords{Cyclops states, rotobreathers, Kuramoto oscillators with inertia, cluster synchronization, higher-harmonic coupling, global bifurcations, period-doubling bifurcations}

\date{\today}
\maketitle

\begin{quotation}
Networks of oscillators often organize into clusters whose members move together while different clusters maintain distinct phases. Cyclops states are a striking example: two synchronized groups coexist with a solitary oscillator, resembling the two ``shoulders'' and single ``eye'' of a Cyclops. Earlier work \cite{munyayev2023cyclops,bolotov2024breathing} showed that higher harmonics in the coupling can make these states dominant in Kuramoto networks with inertia. Here, we ask what happens when the fixed cyclops geometry loses stability. We show that it does not simply disappear. Instead, it gives way to a hierarchy of organized nonstationary motions. Some cyclops states breathe, with the two clusters oscillating around the solitary oscillator. Others rotobreath, with the intercluster phases completing full rotations relative to it. The key surprise is that rotobreathers are not necessarily the next stage in the evolution of breathers. Rather, they can arise through a separate global mechanism, similar in spirit to the emergence of periodic oscillations via a homoclinic bifurcation of a saddle equilibrium in a driven pendulum.  We also identify common breakdown routes of breathing and rotobreathing states, including period-doubling bifurcations  
that create phase-split cyclops states whose intercluster oscillations alternate over two cycles before repeating. Reminiscent of alternating chimera states, these phase-split patterns reveal how identical oscillators can generate self-induced phase disorder 
that can stabilize the formation of non-stationary cyclops states.
Together, these results show how a single solitary oscillator, amplified by inertia and higher-harmonic coupling, can organize the birth and breakdown of complex cluster motion in networks of identical oscillators.

\end{quotation}

\section{Introduction}\label{sec:introduction}
	
Networks of coupled phase oscillators provide a paradigmatic framework for studying collective dynamics in biological and technological systems, including neuronal populations, chemical oscillators, laser arrays, and power grids.\cite{hoppensteadt2012weakly,tinsley2012chimera,ding2019dispersive,dorfler2013synchronization} The classical first-order Kuramoto model~\cite{kuramoto1975self,strogatz2000kuramoto} captures a broad range of transitions from incoherence to collective order.~\cite{acebron,barreto2008synchronization,ott2008low,hong2007entrainment,pikovsky2008partially,maistrenko2004mechanism,dorfler2011critical,skardal2019synchronization} For heterogeneous natural frequencies, these transitions commonly involve partial synchronization and the separation of the population into coherent and incoherent groups.~\cite{martens2009exact,laing2009dynamics} For identical oscillators, the same symmetry can break spontaneously to produce chimera states.~\cite{kuramoto2002coexistence,abrams2004chimera,abrams2008solvable,panaggio2015chimera} Chimeras and related partially coherent patterns have since been found in many oscillator settings,~\cite{omel2013coherence,wolfrum2011spectral,laing2019dynamics,ashwin2015weak,panaggio2016chimera,larger2013virtual,semenova2016coherence,bolotov2016marginal,bolotov2018simple} including chemical oscillators,~\cite{tinsley2012chimera} metronome arrays,~\cite{martens} coupled pendula,~\cite{kapitaniak2014imperfect} pedestrian-induced bridge oscillations,~\cite{belykh2017foot} optical systems,~\cite{hagerstrom2012experimental} and spatially extended media.~\cite{nicolaou2017chimera,smirnov2017chimera}

The Kuramoto model with inertia extends each phase oscillator to a second-order rotator capable of adjusting its instantaneous frequency, as required, for example, in adaptive-frequency models and power-grid dynamics.~\cite{ermentrout1991adaptive,tumash2019stability} This added degree of freedom substantially enriches the collective behavior, producing complex and hysteretic synchronization transitions,~\cite{tanaka1997first,tanaka1997self,ji2014low,munyaev2020analytical,komarov2014synchronization,olmi2014hysteretic} coexisting synchronous clusters,~\cite{belykh2016bistability} chaotic intercluster motion,~\cite{brister2020three} chimeras,~\cite{olmi2015intermittent,maistrenko2017smallest,medvedev2021stability} and solitary states.~\cite{jaros2015chimera,jaros2018solitary,munyayev2022stability} Of particular importance here, inertia allows phase differences between coherent groups to become dynamical variables rather than fixed offsets. In two-population networks, the intercluster phase shift can be governed by a driven-pendulum equation and can therefore remain stationary or execute periodic rotations.~\cite{belykh2016bistability} With three populations, coupled pendulum-like phase differences can display oscillatory, rotatory, mixed-mode, and chaotic dynamics.~\cite{brister2020three} A single oscillator may likewise sustain a stable rotatory phase difference from a coherent cluster.~\cite{munyayev2022stability} This dynamical property of inertia, which allows intercluster phase differences to oscillate and rotate, provides the foundation for the nonstationary cyclops states studied in this paper.

Most Kuramoto-type studies retain only the first sinusoidal term in the Fourier expansion of a general $2\pi$-periodic pairwise interaction function. Although this truncation is analytically convenient, higher Fourier harmonics can be essential for rhythmogenesis. They arise naturally in phase models of neuronal plasticity and learning,~\cite{seliger2002plasticity,niyogi2009learning} electrochemical oscillators,~\cite{kiss2005predicting} and Josephson junctions.~\cite{goldobin2013phase} In Kuramoto networks, higher-mode coupling can create multiple phase-locked states~\cite{komarov2013multiplicity,berner2023synchronization} and induce switching between synchrony clusters.~\cite{skardal2011cluster} Thus, the combination of inertia and higher harmonics provides two complementary ingredients: inertia enables time-dependent intercluster phase dynamics, whereas higher harmonics select and stabilize nontrivial cluster partitions.

The interplay between inertia and higher coupling harmonics led to the discovery of cyclops states in globally coupled networks of identical Kuramoto--Sakaguchi rotators.~\cite{munyayev2023cyclops} In odd-sized networks, a cyclops state consists of a solitary oscillator and two equal coherent clusters, representing the Cyclops’ eye and shoulders, respectively, and
forming a distinguished three-cluster generalized splay state.~\cite{berner2021generalized}
Under repulsive first-harmonic coupling, cyclops states already occupy substantial portions of phase space; remarkably, the addition of a second or third coupling harmonic can make them dominant and, over broad parameter intervals, effectively global attractors.~\cite{munyayev2023cyclops} Cyclops states also persist beyond the perfectly symmetric setting. In particular, recent work revealed a counterintuitive effect: frequency heterogeneity can stabilize stationary cyclops states that are unstable in networks of identical oscillators.~\cite{bolotov2025heterogeneity}

A detailed analysis established the existence and stability conditions for stationary cyclops states and identified two principal mechanisms by which they lose stability.~\cite{bolotov2024breathing} In the first, an Andronov-Hopf bifurcation preserves both coherent clusters while inducing oscillations of their phase differences relative to the solitary oscillator, thereby producing a breathing cyclops state. In the second, a transversal cluster instability disrupts and repeatedly reorganizes the cluster composition, giving rise to switching cyclops dynamics reminiscent of blinking chimeras.~\cite{goldschmidt2019blinking} Rotobreathing states, in which bounded oscillations are accompanied by full rotations of the intercluster phase differences, were also observed; however, their origin, stability domains, and relationship to breathing cyclops states remained unresolved.

In this paper, we provide a systematic analysis of the birth and breakdown of breathing and rotobreathing cyclops states. Our central aim is to determine how nonstationary cluster dynamics is organized after stationary cyclops states with fixed intercluster phase differences lose stability and how the resulting states subsequently destabilize. Specifically, we ask how breathing cyclops states evolve away from their Andronov-Hopf bifurcation onset, whether rotobreathers are merely large-amplitude continuations of breathing states or arise through a distinct mechanism, whether the two families share common routes to instability, and which attractors prevail in inertial Kuramoto networks with higher-mode coupling when the overall interaction is repulsive and complete synchronization is unstable.

Addressing these questions is technically challenging and requires tracking time-dependent cyclops states rather than stationary solutions. Whereas stationary cyclops states appear as fixed points of the reduced intercluster equations, breathing and rotobreathing states correspond to periodic and rotatory trajectories with unknown periods and, in the latter case, nonzero winding numbers. To identify and continue these solutions, we combine a reduction to the three-cluster manifold with Newton-Raphson searches for periodic and rotatory orbits, numerical continuation in the second-harmonic parameters, and Floquet analysis of the full oscillator network. This framework allows us to follow both stable and unstable branches and distinguish bifurcations in intercluster motion from instabilities that split or destroy coherent clusters.

Our analysis uncovers distinct mechanisms underlying the emergence of breathing and rotobreathing cyclops states, as well as common routes for their breakdown. Breathing states undergo period-doubling and cluster-destruction bifurcations, giving rise to phase-split cyclops states, switching dynamics, or multicluster regimes. Rotobreathers, by contrast, predominantly emerge from a separate global bifurcation organized by saddle-cluster states, rather than as a continuation of breathers. Once formed, however, they lose stability through the same two generic secondary scenarios. Attraction-basin statistics further show that breathing and rotobreathing cyclops states are prevalent over broad parameter ranges and serve as organizing structures for repulsive inertial Kuramoto dynamics with higher-mode coupling.

The paper is organized as follows. Section~\ref{sec:model} introduces the network model and defines the stationary, breathing, and rotobreathing cyclops states. Section~\ref{sec:breathing} develops the continuation and Floquet analysis of breathing states and describes their destabilization. Section~\ref{sec:rotobreathing} analyzes rotobreathing states, their global bifurcation boundary, and their secondary instabilities. Section~\ref{sec:persistence} examines attraction basins, persistence in large odd-sized networks, and the contrast with even-sized networks. Section~\ref{sec:conclusion} summarizes the results. The Appendices give the stability analysis of the dominant even-network states and discuss asymmetric cyclops-like configurations for even network sizes.

\section{Network model and possible cyclops dynamics}\label{sec:model}

We consider a globally coupled network of $N$ identical Kuramoto--Sakaguchi phase oscillators~\cite {sakaguchi1986solvable} with inertia and two Fourier harmonics in the coupling function:
\begin{equation}
    \mu \ddot{\theta}_k + \dot{\theta}_k
    = \omega + \frac{1}{N}\sum_{n=1}^{N}\sum_{q=1}^{2}
\varepsilon_q\sin\!\left[q(\theta_n-\theta_k)-\alpha_q\right].
\label{eq:model_def}
\end{equation}
Here, $\theta_k$ denotes the phase of the $k$th oscillator ($k=1,2,\dots,N$), with $-\pi < \theta_k \leq \pi$; $\omega$ is the common natural frequency; $\mu>0$ is the inertia; and $\varepsilon_q$ and $\alpha_q$ are, respectively, the strength and phase lag of the $q$th coupling harmonic. The corresponding pairwise interaction function is
$H(\theta_n - \theta_k) = \sum_{q=1}^2 \varepsilon_q \sin[q(\theta_n - \theta_k) - \alpha_q].$
We set $\varepsilon_1=1$ to fix the coupling scale and choose $\alpha_1\in(\pi/2,\pi)$, so that the first harmonic is repulsive. We also fix the inertia at $\mu=1$, which is sufficiently large to make the second-order oscillator network dynamically distinct from the classical first-order model and to support the oscillatory and rotatory intercluster motions central to the cyclops states studied below. The parameters $\varepsilon_2>0$ and $\alpha_2\in(-\pi,\pi]$ control the second harmonic, which may reinforce or counteract the effect of the first. The sign of
$H'(0)=\varepsilon_1\cos\alpha_1+2\varepsilon_2\cos\alpha_2$
distinguishes the overall attractive regime, $H'(0)>0$, from the overall repulsive regime, $H'(0)<0$, with the fully synchronous state changing its local stability at $H'(0)=0$.

Phase and cluster coherence are characterized by the complex order parameters:
\begin{equation}
    R_l(t)=\frac{1}{N}\sum_{k=1}^{N}e^{il\theta_k}
    =r_l e^{i\psi_l},
    \label{eq:parameter_R}
\end{equation}
where $r_l=|R_l|$. In particular, $r_1=1$ corresponds to complete phase synchronization when $\theta_1 = \theta_2 = \dots = \theta_N$, while $r_2$ measures the degree of two-cluster separation,  and the value $r_2 = 1$ corresponds to a symmetric two-cluster state in a system with an even number of oscillators. With first-harmonic coupling only, generalized splay states $\theta_k = \omega t + \varphi_k$ with constant nonuniform relative phases $\varphi_k \in[-\pi, \pi]$ satisfy $r_1=0$. 

\begin{figure*}[t!]
    \centering
    \includegraphics[width=0.70\linewidth]{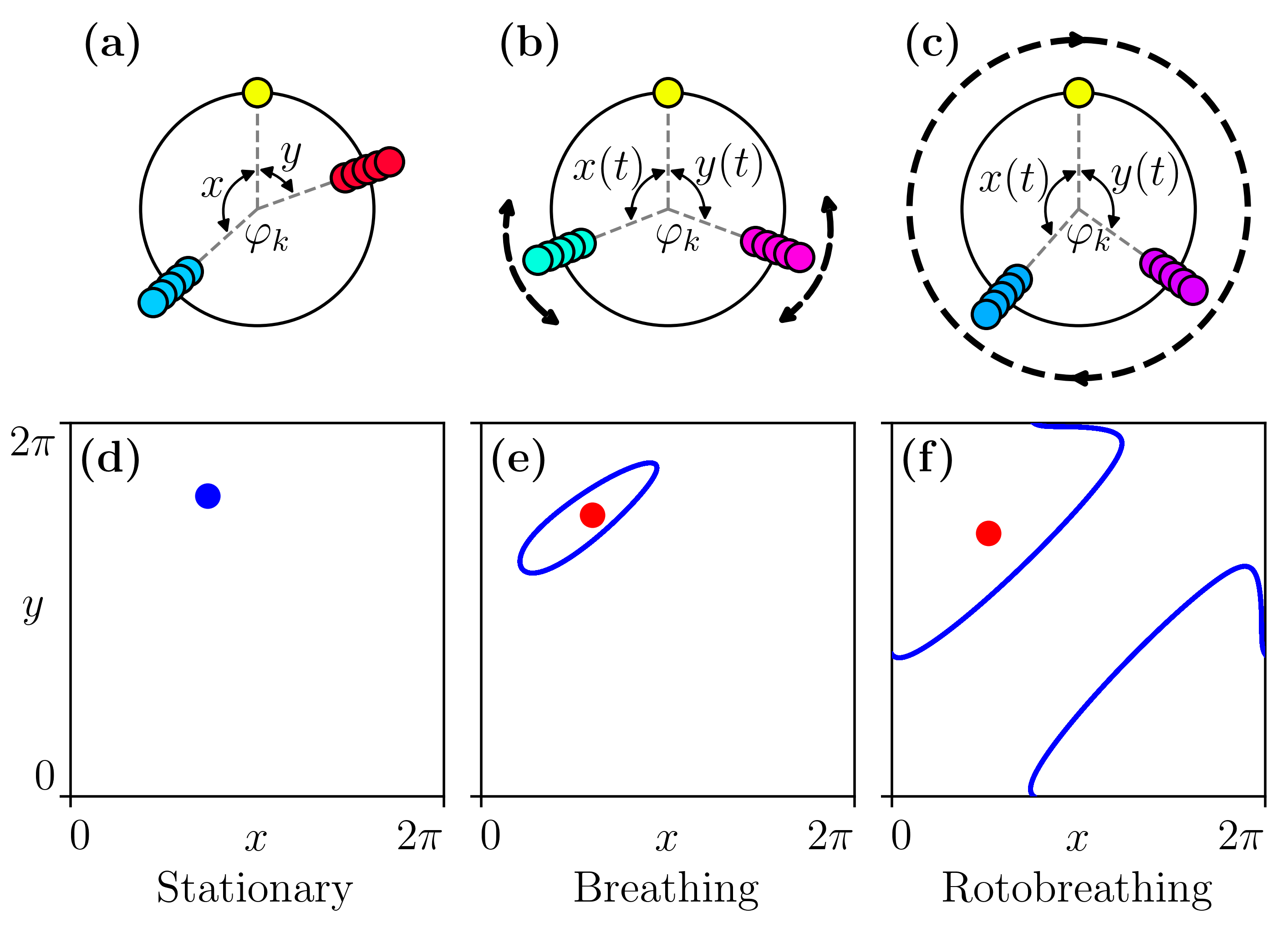}
    \caption{Cyclops states in oscillator-phase and reduced phase-plane representations. (a)-(c) Instantaneous phase configurations relative to the solitary oscillator, $\varphi_k=\theta_k-\theta_M$. The two colored groups are coherent clusters, and the yellow oscillator is solitary. (a) A stationary cyclops state with fixed intercluster phase differences $x$ and $y$. (b) A breathing cyclops state with bounded oscillations $x(t)$ and $y(t)$. (c) A rotobreathing cyclops state with full intercluster rotations; arrows and the dashed circular path indicate the direction and range of the motion. (d)--(f) The corresponding invariant sets of the reduced system~\eqref{eq:xy_dyn} in the $(x,y)$ phase plane: an equilibrium, a closed zero-winding orbit, and a rotatory orbit on the cylindrical phase space, respectively. Blue denotes the stable state or orbit; red dots indicate unstable stationary cyclops equilibria where shown.}
    \label{fig:phase_portraits}
\end{figure*}

For an odd-sized network, $N=2K+1$, the generalized splay state with maximal $r_2$ has the three-cluster phase configuration~\cite{munyayev2023cyclops}
\begin{equation}
\begin{aligned}
    \varphi_1=\cdots=\varphi_K&=\gamma,\qquad \varphi_M=0,\\
    \varphi_{M+1}=\cdots=\varphi_N&=-\gamma,
\end{aligned}
\label{eq:3clast_splay}
\end{equation}
where $M=K+1=(N+1)/2$ and $\gamma=\arccos[1/(1-N)]$. This configuration consists of two equal coherent clusters of size $K=(N-1)/2$ and a solitary oscillator with index $M$, and is termed a cyclops state. The choice $\varphi_M=0$ simply fixes an otherwise arbitrary reference phase. Adding the second coupling harmonic generally shifts the two clusters away from the symmetric positions $\pm\gamma$ to unequal phase offsets $\gamma_{1,2}$ and makes $r_1$ nonzero, while preserving the $K:1:K$ cluster partition. Accordingly, the two-harmonic system~\eqref{eq:model_def} admits asymmetric cyclops solutions on the three-cluster manifold
\begin{equation}
    D(3)=
    \begin{cases}
        \theta_1(t)=\cdots=\theta_K(t)=\Omega t+x,\\
        \theta_M(t)=\Omega t,\\
        \theta_{M+1}(t)=\cdots=\theta_N(t)=\Omega t+y.
    \end{cases}
    \label{eq:3_clast_st}
\end{equation}
Here, the two coherent clusters and the solitary oscillator rotate with a common frequency $\Omega$; its explicit expression is given in Ref.~\cite{bolotov2024breathing}. The constants $x$ and $y$ are the phase differences of the two clusters relative to the $M$th solitary oscillator. Because these phase differences remain fixed, the solution in Eq.~\eqref{eq:3_clast_st} is called a \emph{stationary cyclops state}; in a frame rotating with frequency $\Omega$, it corresponds to an equilibrium of system~\eqref{eq:model_def} (Fig.~\ref{fig:phase_portraits}(a)). Depending on the coupling parameters, the system can support as many as 16 stationary cyclops states with distinct ordered pairs $(x,y)$.~\cite{bolotov2024breathing}

Inertia also permits cyclops states with time-dependent intercluster phase differences. Such states remain on a fixed $K:1:K$ cluster manifold of the form
\begin{equation}
    \begin{array}{l}
        \theta_1(t) = \ldots = \theta_{K}(t) = \psi(t) + x(t), \\
        \theta_M(t) = \psi(t),\\
        \theta_{M+1}(t) = \ldots = \theta_N(t) = \psi(t) + y(t),
    \end{array}
    \label{eq:3_clast_unst}
\end{equation}
where $\psi(t)$ describes the generally nonuniform motion of the solitary oscillator. We call a solution of Eq.~\eqref{eq:3_clast_unst} a \emph{breathing cyclops state} when $|x(t)|<\pi$ and $|y(t)|<\pi$. In this case, the two coherent clusters oscillate relative to the solitary oscillator without undergoing $2\pi$ phase slips (Fig.~\ref{fig:phase_portraits}(b)). A \emph{rotobreathing cyclops state} occurs when $|x(t)|>\pi$ or $|y(t)|>\pi$ during the motion. Equivalently, at least one of the phase differences has a nonzero winding number, so that the oscillatory modulation is accompanied by full rotations relative to the solitary oscillator (Fig.~\ref{fig:phase_portraits}(c)).
A {\it switching cyclops} state completes the classification considered here. Unlike the stationary, breathing, and rotobreathing states, it does not remain on a fixed $K:1:K$ cluster partition. Instead, one coherent cluster repeatedly disintegrates and reforms with a reshuffled composition, thereby selecting a new solitary oscillator and a new cyclops configuration.

Previous work identified two primary routes by which a stationary cyclops state loses stability.~\cite{bolotov2024breathing} An instability within the three-cluster manifold produces time-dependent $x(t)$ and $y(t)$ through an Andronov--Hopf bifurcation and gives rise to breathing dynamics. A transverse instability disrupts one of the coherent clusters, while the remaining stable directions of the saddle-type dynamics keep the trajectory near cyclops configurations and enable their repeated reformation, producing switching dynamics. In the present paper, these mechanisms serve as starting points for studying the subsequent breakdown of breathing cyclops states and the distinct birth and breakdown mechanisms of rotobreathing cyclops states.

Toward this goal, we subtract the equation for the solitary oscillator $\theta_M$ in \eqref{eq:model_def}
from the equations for the two coherent clusters and eliminate the common phase $\psi(t)$. The natural frequency $\omega$ also cancels from the phase-difference dynamics, so it may be set to zero in the reduced equations without loss of generality. We obtain
\begin{widetext}
\begin{equation}
\begin{aligned}
    \mu\ddot{x}+\dot{x}
    &=\sum_{q=1}^{2}\frac{\varepsilon_q}{N}\Big[
    \sin\alpha_q-\sin(qx+\alpha_q)
    -K\big(\sin(qx-\alpha_q)+\sin(qy-\alpha_q)
    +\sin\alpha_q+\sin(q(x-y)+\alpha_q)\big)\Big],\\
    \mu\ddot{y}+\dot{y}
    &=\sum_{q=1}^{2}\frac{\varepsilon_q}{N}\Big[
    \sin\alpha_q-\sin(qy+\alpha_q)
    -K\big(\sin(qx-\alpha_q)+\sin(qy-\alpha_q)
    +\sin\alpha_q+\sin(q(y-x)+\alpha_q)\big)\Big].
\end{aligned}
\label{eq:xy_dyn}
\end{equation}
\end{widetext}
Equilibria of Eq.~\eqref{eq:xy_dyn} represent stationary cyclops states (Fig.~\ref{fig:phase_portraits}(d)), closed periodic orbits with zero winding numbers represent breathing states (Fig.~\ref{fig:phase_portraits}(e)), and rotatory orbits with nonzero winding numbers represent rotobreathing states (Fig.~\ref{fig:phase_portraits}(f)).

The four-dimensional phase-difference system~\eqref{eq:xy_dyn} has a similar basic structure as the intercluster equations derived for three-population inertial Kuramoto networks in Eq.~(6) of Ref.~\cite{brister2020three}. In that setting, a change of variables casts the dynamics as two coupled driven-pendulum-like equations capable of supporting periodic, rotatory, mixed-mode, and chaotic intercluster motion.\cite{brister2020three} The same dynamical ingredients are present here: inertia makes each phase difference second order, while the two coupling harmonics generate nonlinear torques and coupling between the phase differences, thereby potentially enabling mixed-mode and chaotic phase-offset dynamics. Analytical bounds based on auxiliary-system methods, similar to those developed for related inertial Kuramoto systems,~\cite{brister2020three,barabash2021partial} may therefore be possible. A full qualitative treatment of Eq.~\eqref{eq:xy_dyn} is beyond the scope of the present work and is left for future study. Here, we instead locate and continue its periodic and rotatory solutions and determine their stability in the full oscillator network.

\section{Breathing cyclops states: birth and breakdown}\label{sec:breathing}

Breathing cyclops states are zero-winding periodic orbits of the reduced system~\eqref{eq:xy_dyn}. Subsections~\ref{sec:breathing:found} and \ref{sec:breathing:stab} describe the numerical procedures used to locate these orbits and test their full-network stability. Readers interested primarily in the bifurcation results may proceed directly to Sec.~\ref{sec:breathing:birth} without loss of continuity.

\subsection{Numerical identification and continuation of breathing orbits}\label{sec:breathing:found}

Let
$\mathbf{X}=(x,\dot{x},y,\dot{y})$
and denote the flow of Eq.~\eqref{eq:xy_dyn} by $\mathbf{F}(t,\mathbf{X}_0)$. A breathing orbit with period $T$ satisfies
$\mathbf{F}(T,\mathbf{X}_0^{(p)})=\mathbf{X}_0^{(p)}$.
Equivalently, it is a fixed point of the time-$T$ map
\begin{equation}
    \mathbf{P}:\mathbf{X}_n\mapsto\mathbf{X}_{n+1},
    \qquad \mathbf{X}_{n+1}=\mathbf{F}(T,\mathbf{X}_n).
\end{equation}
We therefore solve the nonlinear periodicity condition
\begin{equation}
    \mathbf{G}(T,\mathbf{X}_0)
    =\mathbf{F}(T,\mathbf{X}_0)-\mathbf{X}_0=\mathbf{0}.
    \label{eq:exist_syst_br}
\end{equation}
Because every point on a periodic orbit satisfies Eq.~\eqref{eq:exist_syst_br}, a scalar phase condition is added to remove the neutral time-translation freedom; in the computations, we fix one coordinate of the initial point, for example $x(0)$. The resulting system is solved by a Newton--Raphson iteration for the unknown period $T$ and initial state $\mathbf{X}_0^{(p)}$. Initial guesses are obtained from direct simulations and are then transferred between nearby parameter values during continuation in $(\alpha_2,\varepsilon_2)$. This procedure follows both stable and unstable periodic branches.
~\\
\subsection{Floquet test of breathing-orbit stability in the full network}\label{sec:breathing:stab}

A periodic orbit of the reduced system must remain stable not only within the three-cluster manifold but also against perturbations that split either coherent cluster. To include these transverse directions, we rewrite Eq.~\eqref{eq:model_def} as
\begin{equation}
    \mu\ddot{\theta}_k+\dot{\theta}_k
    =\frac{1}{N}\sum_{n=1}^{N}H(\theta_n-\theta_k)
    \label{eq:theta_H}
\end{equation}
and introduce the phase differences of all cluster oscillators relative to the solitary oscillator,
\begin{equation}
    x_i=\theta_i-\theta_M,
    \qquad y_i=\theta_{M+i}-\theta_M,
    \qquad i=1,\ldots,K.
    \label{eq_xi_yi}
\end{equation}
They satisfy
\begin{widetext}
\begin{equation}
\begin{aligned}
    \mu\ddot{x}_i+\dot{x}_i
    &=\frac{1}{N}\Bigg[H(-x_i)
      +\sum_{j=1}^{K}H(x_j-x_i)
      +\sum_{j=1}^{K}H(y_j-x_i)
      -H(0)-\sum_{j=1}^{K}H(x_j)-\sum_{j=1}^{K}H(y_j)\Bigg],\\
    \mu\ddot{y}_i+\dot{y}_i
    &=\frac{1}{N}\Bigg[H(-y_i)
      +\sum_{j=1}^{K}H(x_j-y_i)
      +\sum_{j=1}^{K}H(y_j-y_i)
      -H(0)-\sum_{j=1}^{K}H(x_j)-\sum_{j=1}^{K}H(y_j)\Bigg].
\end{aligned}
\label{eq:phase_diverg}
\end{equation}
\end{widetext}
For a cyclops orbit, $x_i(t)=x(t)$ and $y_i(t)=y(t)$. We set
$x_i=x+\delta x_i$ and $y_i=y+\delta y_i$ and linearize Eq.~\eqref{eq:phase_diverg} to obtain
\begin{equation}
\begin{aligned}
    \mu\delta\ddot{x}_i+\delta\dot{x}_i
    &=\frac{1}{N}\left[-A_x\delta x_i
    +B_x\sum_{j=1}^{K}\delta x_j
    +C_x\sum_{j=1}^{K}\delta y_j\right],\\
    \mu\delta\ddot{y}_i+\delta\dot{y}_i
    &=\frac{1}{N}\left[-A_y\delta y_i
    +B_y\sum_{j=1}^{K}\delta y_j
    +C_y\sum_{j=1}^{K}\delta x_j\right],
\end{aligned}
\label{eq:after_ABC}
\end{equation}
where
\begin{align*}
    A_x(x,y)&=H'(-x)+KH'(0)+KH'(y-x),\\
    B_x(x)&=H'(0)-H'(x),\\
    C_x(x,y)&=H'(y-x)-H'(y),\\
    A_y(x,y)&=H'(-y)+KH'(0)+KH'(x-y),\\
    B_y(y)&=H'(0)-H'(y),\\
    C_y(x,y)&=H'(x-y)-H'(x).
\end{align*}

Writing the perturbation vector as
\[
\delta\mathbf{Z}=(\delta x_1,\delta\dot{x}_1,\ldots,\delta x_K,\delta\dot{x}_K,
\delta y_1,\delta\dot{y}_1,\ldots,\delta y_K,\delta\dot{y}_K)^{\mathsf T},
\]
we obtain the $4K$-dimensional periodic variational system
\begin{equation}
    \delta\dot{\mathbf{Z}}=\mathbf{J}(t)\delta\mathbf{Z}.
    \label{eq:floquet_system}
\end{equation}
Along the periodic orbit $\mathbf{F}_p(t,\mathbf{X}_0^{(p)})$,
\[
\mathbf{J}(t)=
\begin{bmatrix}
\mathbf{J}_{11}(x,y)&\mathbf{J}_{12}(x,y)\\
\mathbf{J}_{21}(x,y)&\mathbf{J}_{22}(x,y)
\end{bmatrix},
\]
with
\begin{align*}
\mathbf{J}_{11}&=\mathbf{I}_K\otimes\mathbf{D}_1
 +(\mathbf{E}_K-\mathbf{I}_K)\otimes\mathbf{D}_2,\\
\mathbf{J}_{12}&=\mathbf{E}_K\otimes\mathbf{D}_3,\\
\mathbf{J}_{21}&=\mathbf{E}_K\otimes\mathbf{D}_4,\\
\mathbf{J}_{22}&=\mathbf{I}_K\otimes\mathbf{D}_5
 +(\mathbf{E}_K-\mathbf{I}_K)\otimes\mathbf{D}_6.
\end{align*}
where $\mathbf{I}_K$ is the identity matrix, $\mathbf{E}_K$ is the all-ones matrix, $\otimes$ denotes the Kronecker product, and
\begin{align*}
\mathbf{D}_1(x,y)&=
\begin{bmatrix}0&1\\(\mu N)^{-1}[B_x(x)-A_x(x,y)]&-\mu^{-1}\end{bmatrix},\\
\mathbf{D}_2(x)&=
\begin{bmatrix}0&0\\(\mu N)^{-1}B_x(x)&0\end{bmatrix},\\
\mathbf{D}_3(x,y)&=
\begin{bmatrix}0&0\\(\mu N)^{-1}C_x(x,y)&0\end{bmatrix},\\
\mathbf{D}_4(x,y)&=
\begin{bmatrix}0&0\\(\mu N)^{-1}C_y(x,y)&0\end{bmatrix},\\
\mathbf{D}_5(x,y)&=
\begin{bmatrix}0&1\\(\mu N)^{-1}[B_y(y)-A_y(x,y)]&-\mu^{-1}\end{bmatrix},\\
\mathbf{D}_6(y)&=
\begin{bmatrix}0&0\\(\mu N)^{-1}B_y(y)&0\end{bmatrix}.
\end{align*}

The fundamental matrix $\boldsymbol{\Phi}(t)$ satisfies
$\dot{\boldsymbol{\Phi}}=\mathbf{J}(t)\boldsymbol{\Phi}$ and
$\boldsymbol{\Phi}(0)=\mathbf{I}_{4K}$. The monodromy matrix advances perturbations through one period,
\begin{equation}
    \delta\mathbf{Z}(T)=\boldsymbol{\Phi}(T)\delta\mathbf{Z}(0).
    \label{eq:monodromy_action}
\end{equation}
Its eigenvalues $\lambda_j$ are the Floquet multipliers. One multiplier is equal to unity because of time-translation invariance. The orbit is linearly stable when every remaining multiplier lies inside the unit circle. We denote the leading nontrivial multiplier by
\begin{equation}
    \lambda_{\max}=\lambda_{i^*},
    \qquad i^*=\underset{j\in\mathcal{I}}{\operatorname{argmax}}\,|\lambda_j|,
    \label{eq:lambda_max}
\end{equation}
where $\mathcal{I}$ excludes the neutral multiplier. Crossings of $\lambda_{\max}$ through $-1$ and $+1$ organize the two breakdown scenarios described below.
\subsection{Birth and existence boundaries of breathing cyclops states}
\label{sec:breathing:birth}

\begin{figure*}[t!]
    \centering
    \includegraphics[width=0.80\linewidth]{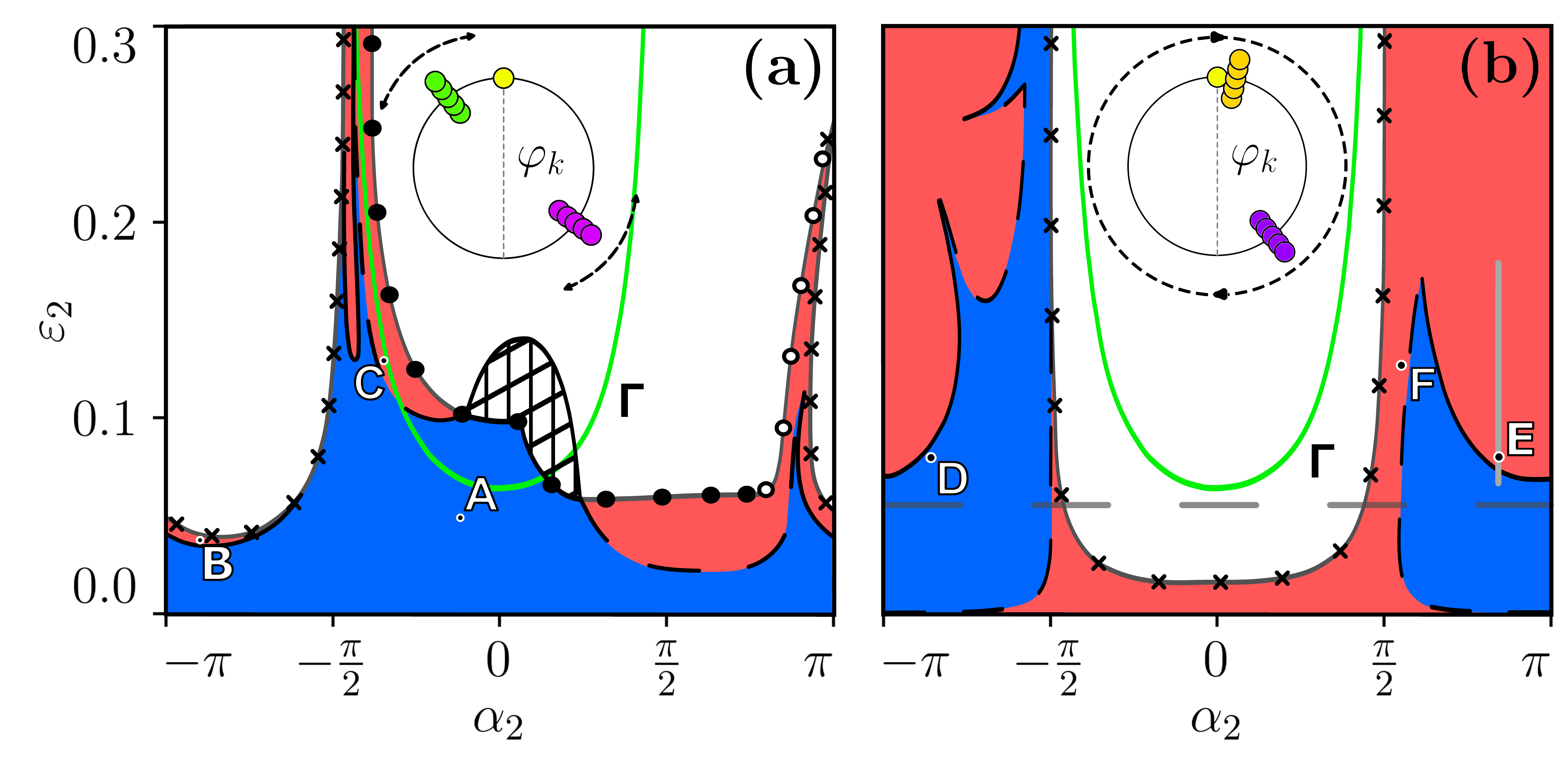}
    \caption{Existence and stability domains of breathing and rotobreathing cyclops states in the $(\alpha_2,\varepsilon_2)$ parameter plane.
    (a) Breathing cyclops states.
    (b) Rotobreathing cyclops states.
    Blue and red indicate stable and unstable periodic or rotatory solutions, respectively. The solid black curves mark stability loss through a Floquet-multiplier crossing at $\lambda_{\max}=-1$, whereas the dashed black curves mark crossings at $\lambda_{\max}=+1$. In panel (a), the gray existence-boundary curves are distinguished by black markers: filled circles indicate the Andronov--Hopf bifurcation at which a breathing orbit is born from a stationary cyclops state; open circles indicate saddle-node bifurcations of periodic orbits; and crosses indicate the global boundary at which a breathing orbit merges with a heteroclinic-contour-like structure formed by the two symmetry-related saddle two-cluster states $(K+1):K$ and $K:(K+1)$. In panel (b), the gray curve with black crosses marks the corresponding global merging boundary for rotatory orbits. The double-hatched regions indicate the existence of stationary cyclops states. The green solid curve $\Gamma$ denotes $H'(0)=0$ and separates the overall repulsive and attractive coupling regimes. The insets illustrate the relative cluster motion, with $\varphi_k=\theta_k-\theta_M$ measured from the solitary oscillator; dashed lines indicate the range of motion and arrows show its direction. Points $A$--$F$ correspond to the representative states examined in Figs.~\ref{fig:br_stable_A_full}--\ref{fig:rb_unstable_C_full}. Periodic and rotatory solutions are identified using the procedures described in Secs.~\ref{sec:breathing:found} and~\ref{sec:rotobreathing:found}, and their stability is determined from the corresponding Floquet multipliers. Parameters: $N=11$, $\mu=1$, $\varepsilon_1=1$, and $\alpha_1=1.7$.}
    \label{fig:br_stable_area}
\end{figure*}

To determine the existence and stability domains of breathing cyclops states in the $(\alpha_2,\varepsilon_2)$ parameter plane, we continue the periodic solutions obtained from Eq.~\eqref{eq:exist_syst_br}. At each parameter point, the periodic orbit is located numerically, and its stability in the full oscillator network is determined from the Floquet multipliers defined through Eq.~\eqref{eq:monodromy_action}. Figure~\ref{fig:br_stable_area}(a) shows the resulting map for $N=11$, $\mu=1$, $\varepsilon_1=1$, and $\alpha_1=1.7$. The value $\alpha_1=1.7$ is chosen because it supports a broad variety of cyclops dynamics.~\cite{bolotov2024breathing} Blue regions indicate stable breathing cyclops states, whereas red regions indicate breathing orbits that exist but are unstable.

Two distinct classes of boundaries appear in Fig.~\ref{fig:br_stable_area}(a). The gray curves with black markers delimit the existence of the breathing-orbit branches: filled circles mark an Andronov--Hopf bifurcation, open circles mark a saddle-node bifurcation of periodic orbits, and crosses mark a global heteroclinic-type boundary. The solid and dashed black curves lie within the existence domain and mark destabilization of an existing breathing orbit through Floquet-multiplier crossings at $\lambda_{\max}=-1$ and $\lambda_{\max}=+1$, respectively. Thus, a breathing orbit continues as an unstable periodic solution after crossing a black curve, whereas it ceases to exist at a gray existence boundary. The dynamics resulting from the two stability losses are considered in Sec.~\ref{sec:breathing:breakdown}.

The breathing branch is born at the Andronov--Hopf boundary, shown by the gray curve with black filled circles. As this boundary is approached along the periodic branch, the amplitude of the intercluster oscillations tends to zero, and the breathing orbit collapses onto the equilibrium corresponding to a stationary cyclops state. The stationary and breathing states therefore meet continuously at this boundary.

The gray curve with black open circles marks a saddle-node bifurcation of periodic orbits. Here, two breathing cycles coalesce and disappear while retaining a finite oscillation amplitude. In contrast to the Andronov--Hopf boundary, the periodic solution does not approach a stationary cyclops state before the branch terminates.

The gray curve with black crosses marks a global bifurcation at which the breathing limit cycle merges with a heteroclinic-contour-like structure formed by saddle two-cluster states. The numerical continuation and phase-space geometry provide the following interpretation of this boundary. As it is approached, one of the intercluster phase differences, $x(t)$ or $y(t)$, remains close to zero for an increasingly long portion of each oscillation. When $x=0$, the first coherent cluster temporarily coincides with the solitary oscillator and forms a $(K+1):K$ two-cluster configuration. Similarly, $y=0$ corresponds to the symmetry-related configuration $K:(K+1)$. These configurations act as saddle cluster states.

Near the global boundary, the breathing trajectory alternates between the neighborhoods of these two saddle configurations and spends progressively longer intervals close to each one. Consequently, the period of the intercluster oscillations grows rapidly. At the limiting parameter value, the periodic orbit merges with a contour-like connection between the saddle states and terminates. The approach to the two symmetry-related saddle configurations and the accompanying increase in period are characteristic signatures of a global heteroclinic bifurcation. Because proving the existence of an exact saddle connection is beyond the scope of the present study, we refer to the limiting structure as heteroclinic-contour-like.

The three gray boundaries therefore correspond to qualitatively different birth or termination mechanisms. At the Andronov--Hopf boundary, a breathing orbit is born from a stationary cyclops state with vanishing oscillation amplitude. At the saddle-node boundary, two finite-amplitude breathing orbits merge and disappear. At the global boundary, the orbit approaches the saddle two-cluster configurations, its period increases without bound, and the branch terminates by merging with the heteroclinic-contour-like structure.

\begin{figure}[t!]
    \centering
    \includegraphics[width=1.0\linewidth]{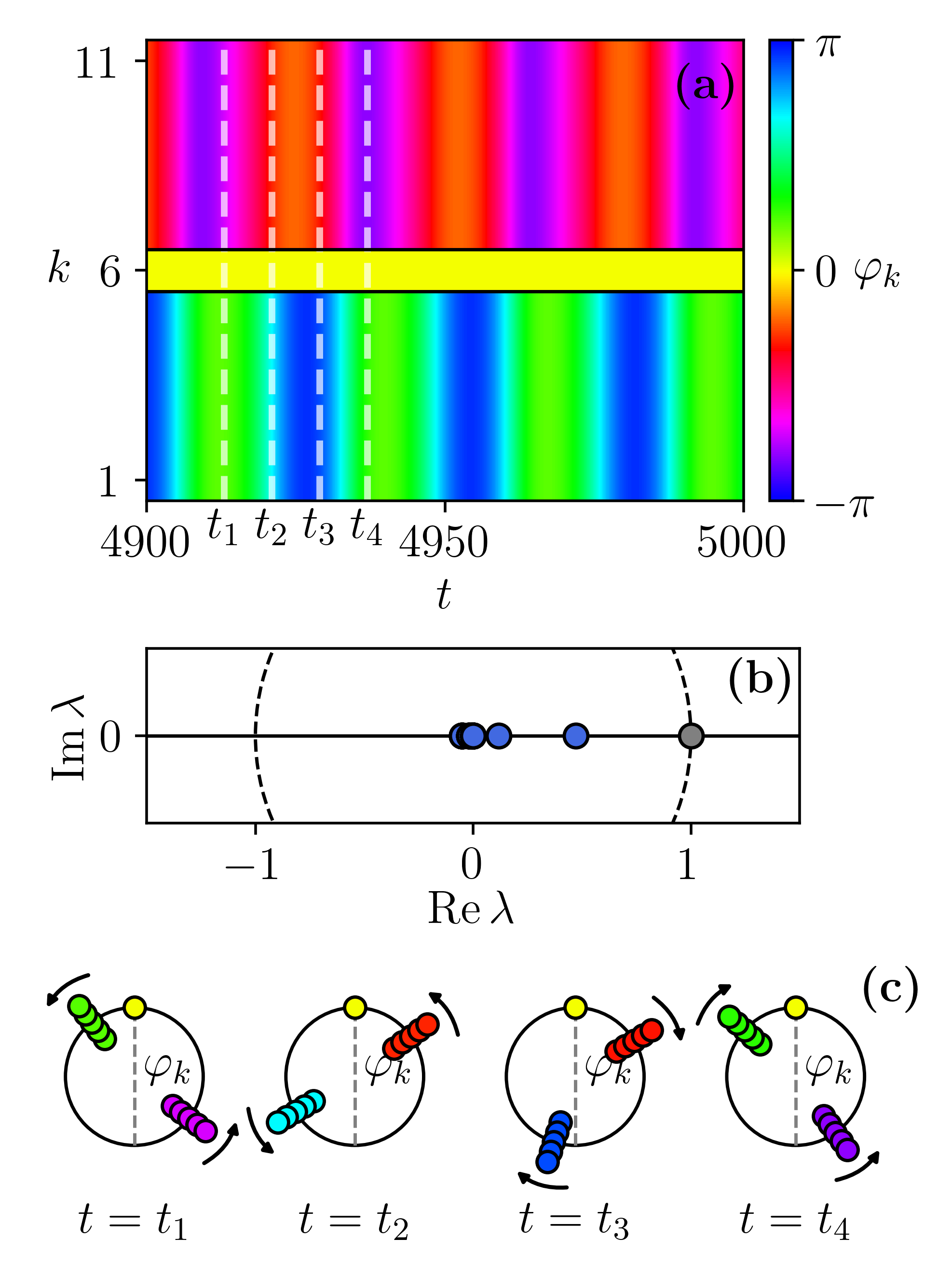}
    \caption{Stable breathing cyclops state at point $A$ in Fig.~\ref{fig:br_stable_area}(a).
    (a) Phase differences $\varphi_k=\theta_k-\theta_6$ relative to the solitary oscillator. The two coherent clusters remain internally synchronized and undergo bounded periodic oscillations without $2\pi$ phase slips. Vertical dashed lines mark the times used for the phase snapshots in panel (c).
    (b) Floquet multipliers of the breathing orbit. All nontrivial multipliers, shown in blue, lie inside the unit circle; the gray multiplier $\lambda_{4K}=1$ is the neutral multiplier associated with time-translation invariance of the periodic orbit.
    (c) Instantaneous phase configurations at $t_1=4913$, $t_2=4921$, $t_3=4929$, and $t_4=4937$. Arrows indicate the direction of the cluster motion relative to the solitary oscillator. The periodic orbit is identified using the procedure in Sec.~\ref{sec:breathing:found}, and its stability is determined from the monodromy matrix in Eq.~\eqref{eq:monodromy_action}. Parameters: $N=11$, $\mu=1$, $\varepsilon_1=1$, $\alpha_1=1.7$, $\varepsilon_2=0.049$, and $\alpha_2=-0.377$.}
    \label{fig:br_stable_A_full}
\end{figure}

As a representative stable breathing cyclops state, we consider point $A$ in Fig.~\ref{fig:br_stable_area}(a), with $\varepsilon_2=0.049$ and $\alpha_2=-0.377$. This point lies inside the blue stability region. The corresponding trajectory is a stable limit cycle in the $(x,y)$ phase plane (Fig.~\ref{fig:phase_portraits}(e)). Direct numerical integration of the full system~\eqref{eq:model_def}, initialized near this orbit, produces the dynamics shown in Fig.~\ref{fig:br_stable_A_full}. The phase differences in Fig.~\ref{fig:br_stable_A_full}(a) remain bounded and periodic, with $|x(t)|<\pi$ and $|y(t)|<\pi$, while the oscillators within each of the two coherent clusters remain perfectly synchronized. All nontrivial Floquet multipliers lie inside the unit circle (Fig.~\ref{fig:br_stable_A_full}(b)), confirming the stability of the breathing state. The snapshots in Fig.~\ref{fig:br_stable_A_full}(c) show the two coherent clusters oscillating relative to the solitary oscillator without undergoing phase slips.

\subsection{Destabilization routes of breathing cyclops states}
\label{sec:breathing:breakdown}

Within their existence domain, breathing cyclops states lose stability through two generic real-multiplier bifurcations. In both cases, the parent period-$T$ breathing orbit continues beyond the stability boundary as an unstable periodic solution. The post-bifurcation outcome, however, depends on whether the leading Floquet multiplier crosses the unit circle through $-1$ or $+1$.

\begin{figure*}[t!]
    \centering
    \includegraphics[width=0.70\linewidth]{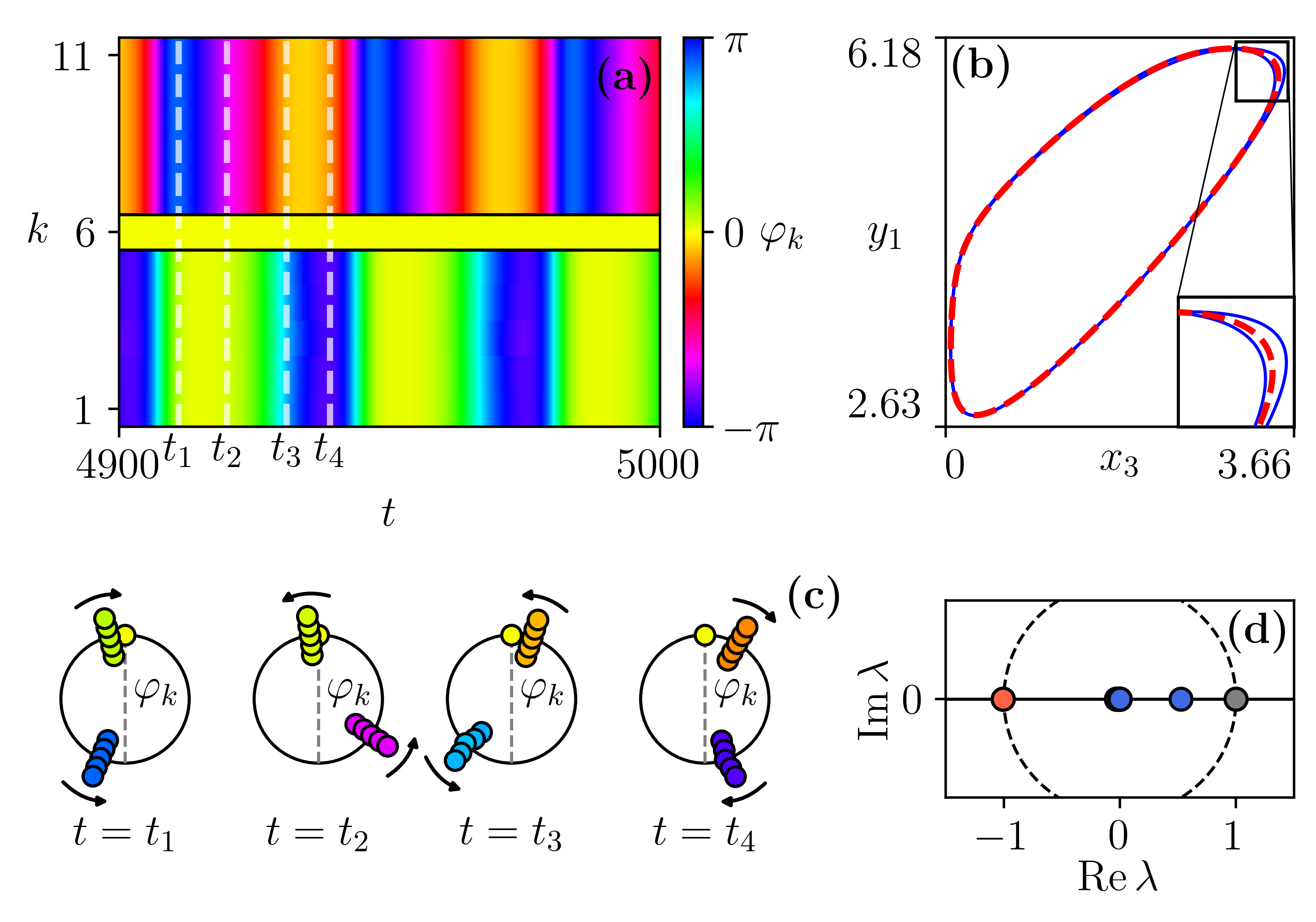}
    \caption{Period-doubling destabilization of a breathing cyclops state and formation of a phase-split breathing cyclops state at point $B$ in Fig.~\ref{fig:br_stable_area}(a).
    (a) Phase differences $\varphi_k=\theta_k-\theta_6$ relative to the solitary oscillator. The oscillators originating from the parent coherent clusters remain frequency locked, but a weak persistent internal phase splitting develops. Vertical dashed lines mark the snapshot times used in panel (c).
    (b) Projection onto the $(x_3,y_1)$ plane. The red dashed curve is the unstable period-$T$ breathing orbit obtained by Newton--Raphson continuation, whereas the blue curve is the attracting period-$2T$ descendant. The inset resolves the small separation between the two trajectories near the period-doubling boundary.
    (c) Instantaneous phase configurations at $t_1=4911$, $t_2=4920$, $t_3=4931$, and $t_4=4939$. The internal phase splitting is most visible at $t_3$ and $t_4$.
    (d) Floquet multipliers of the parent period-$T$ orbit. The red multiplier has crossed $-1$, the blue multipliers remain inside the unit circle, and the gray multiplier at $1$ is neutral. Parameters: $N=11$, $\mu=1$, $\varepsilon_1=1$, $\alpha_1=1.7$, $\varepsilon_2=0.0346$, and $\alpha_2=-2.765$.}
    \label{fig:br_unstable_B_full}
\end{figure*}

At point $B$ in Fig.~\ref{fig:br_stable_area}(a), with $\varepsilon_2=0.0346$ and $\alpha_2=-2.765$, the leading nontrivial Floquet multiplier crosses $-1$. The parent breathing orbit therefore undergoes a period-doubling bifurcation. In Fig.~\ref{fig:br_unstable_B_full}(b), the red dashed curve shows the unstable period-$T$ orbit obtained by Newton--Raphson continuation, whereas the blue curve shows the attracting descendant reached after a small perturbation. The descendant does not return to the same phase configuration after one period $T$ of the parent breather. Instead, its intercluster motion and internal phase arrangement alternate over two cycles and repeat only after $2T$.

The period-doubling bifurcation destabilizes the parent breathing orbit but does not destroy the cyclops organization. The distinguished solitary oscillator persists, and the oscillators originating from each parent coherent cluster remain frequency locked. Exact phase synchronization within one of the clusters is nevertheless lost: its oscillators acquire persistent relative phase offsets and may divide into smaller phase-synchronized subclusters. We call the resulting regime a \emph{phase-split breathing cyclops state}. Thus, the period-doubling route transforms the original breathing state into a more complex cyclops configuration rather than causing its complete breakdown.

The phase splitting at point $B$ is weak because the leading multiplier has crossed the unit circle only slightly. Accordingly, the unstable parent orbit and the attracting period-doubled orbit remain close in Fig.~\ref{fig:br_unstable_B_full}(b), and their separation is resolved more clearly in the inset. The internal phase spread is also only weakly visible in the phase-difference diagram in Fig.~\ref{fig:br_unstable_B_full}(a), but it can be distinguished in the snapshots at $t_3$ and $t_4$ in Fig.~\ref{fig:br_unstable_B_full}(c). The Floquet spectrum of the parent orbit (Fig.~\ref{fig:br_unstable_B_full}(d)) confirms that the loss of stability occurs through a real multiplier crossing at $-1$.

Along the parameter path considered here, moving farther from the period-doubling boundary eventually causes trajectories to approach a stable rotobreathing attractor. This change of the observed attractor should not be interpreted as a smooth continuation of the breathing branch into the rotobreathing branch. As shown in Sec.~\ref{sec:rotobreathing:birth}, rotobreathing states possess a distinct global origin.

\begin{figure*}[t!]
    \centering
    \includegraphics[width=0.70\linewidth]{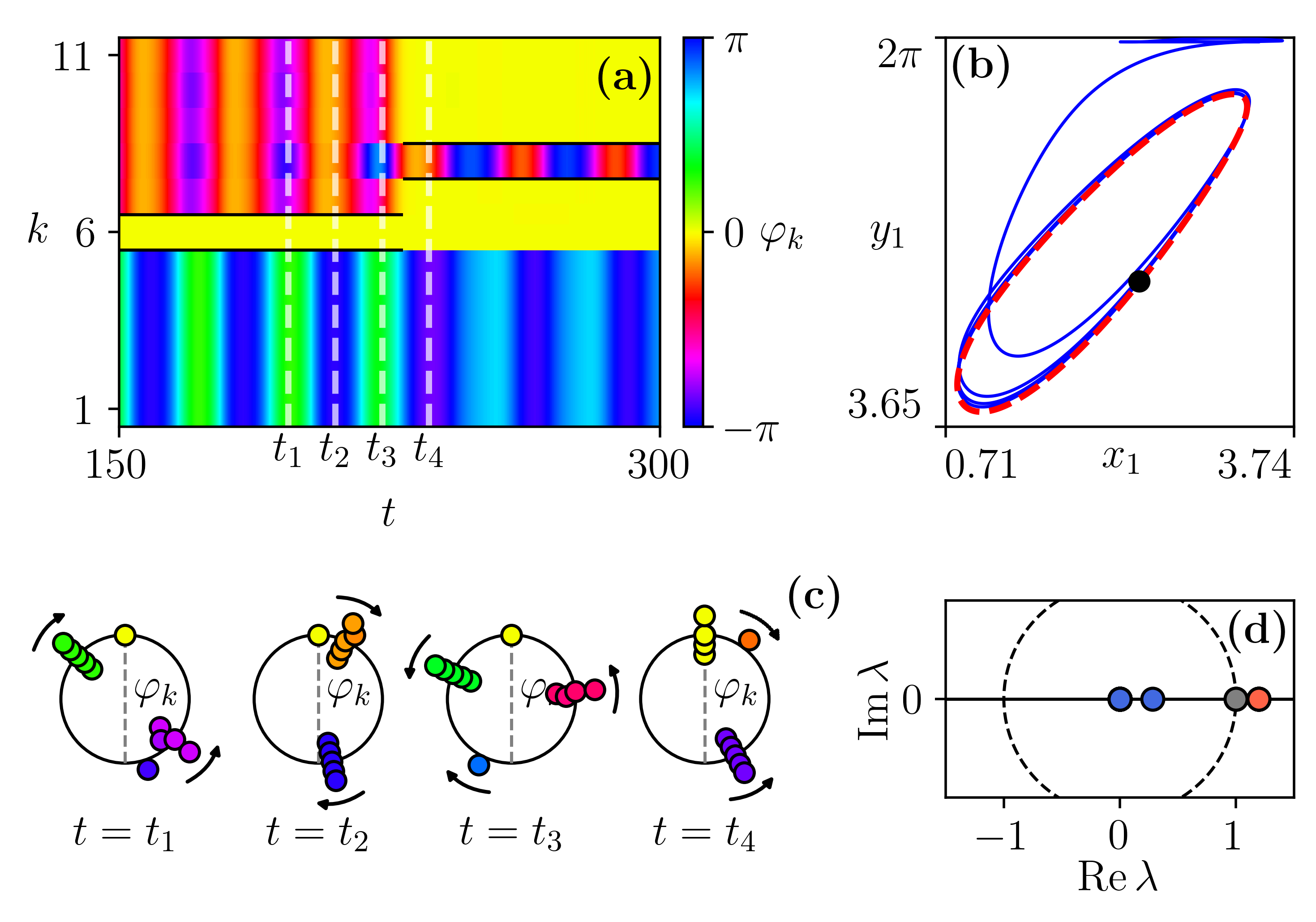}
    \caption{Cluster-destruction breakdown of a breathing cyclops state and transition to switching dynamics at point $C$ in Fig.~\ref{fig:br_stable_area}(a).
    (a) Phase differences relative to the original solitary oscillator show recurrent changes in cluster membership. Vertical dashed lines mark the snapshot times used in panel (c).
    (b) Projection of the reduced dynamics. The red dashed curve is the unstable breathing orbit, the black dot marks the initial perturbation, and the blue trajectory departs from the orbit toward the switching attractor.
    (c) Instantaneous phase configurations at $t_1=197$, $t_2=210$, $t_3=223$, and $t_4=236$. The original solitary oscillator joins a coherent cluster, while another oscillator separates and assumes the solitary role.
    (d) Floquet multipliers of the parent breathing orbit. The red multiplier has crossed $+1$, the blue multipliers remain inside the unit circle, and the gray multiplier at $1$ is neutral. Parameters: $N=11$, $\mu=1$, $\varepsilon_1=1$, $\alpha_1=1.7$, $\varepsilon_2=0.13$, and $\alpha_2=-1.169$.}
    \label{fig:br_unstable_C_full}
\end{figure*}

A second destabilization route occurs at point $C$ in Fig.~\ref{fig:br_stable_area}(a), with $\varepsilon_2=0.13$ and $\alpha_2=-1.169$. Here, the leading Floquet multiplier crosses $+1$. Unlike the period-doubling instability, the associated perturbation destroys the coherence of one of the parent clusters. The red dashed curve in Fig.~\ref{fig:br_unstable_C_full}(b) represents the unstable breathing orbit, while the blue trajectory, initiated by the perturbation marked by the black dot, departs from its neighborhood toward a switching attractor.

The corresponding full-network dynamics are shown in Figs.~\ref{fig:br_unstable_C_full}(a) and~\ref{fig:br_unstable_C_full}(c). The original solitary oscillator approaches and joins one of the coherent clusters. At approximately the same stage, an oscillator from that cluster separates and becomes the new solitary unit. This process repeats, continually reassigning the solitary oscillator and reorganizing the cluster membership. The resulting regime is a switching cyclops state.~\cite{bolotov2024breathing} Depending on the parameters, the same cluster-destruction bifurcation may instead lead to more general multicluster dynamics.

In contrast to the period-doubling route, this $+1$ instability represents a genuine breakdown of the parent breathing cyclops state because its original coherent-cluster composition is destroyed. The Floquet spectrum in Fig.~\ref{fig:br_unstable_C_full}(d) confirms that the parent orbit loses stability through a real multiplier crossing at $+1$.

Breathing cyclops states involve bounded oscillations of the intercluster phase differences, whereas inertia also permits these phase differences to wind around the circle. Such rotatory dynamics introduce a topologically distinct form of collective motion and, in the four-dimensional reduced phase-difference system, can interact with oscillatory modes to generate mixed-mode and potentially chaotic phase-offset dynamics.~\cite{belykh2016bistability,brister2020three,munyayev2022stability} It is therefore important to determine whether rotobreathing cyclops states arise as a continuation of breathing motion or through an independent mechanism, and whether the two families share common routes to instability.

\section{Rotobreathing cyclops states: emergence and breakdown}
\label{sec:rotobreathing}

Rotobreathing cyclops states correspond to periodic motion on the cylindrical phase space with a nonzero winding number in at least one intercluster phase difference. Unlike breathing states, their trajectories are not closed in the unwrapped $(x,y)$ phase plane but become periodic after the accumulated $2\pi$ rotations are removed. 

\subsection{Numerical identification and Floquet test of rotatory orbits}
\label{sec:rotobreathing:found}

Let $\mathbf{F}_r(t,\mathbf{X}_0^{(r)})$ denote a rotatory solution of Eq.~\eqref{eq:xy_dyn}. In addition to its period $T$, the solution is characterized by integer winding numbers $w_x$ and $w_y$ of the phase differences $x(t)$ and $y(t)$:
\begin{equation}
    \mathbf{F}_r(t+T,\mathbf{X}_0^{(r)})
    =\mathbf{F}_r(t,\mathbf{X}_0^{(r)})
    +(2\pi w_x,0,2\pi w_y,0).
    \label{eq:rb_period}
\end{equation}
Thus, over one period, the phase differences may accumulate integer multiples of $2\pi$, while their velocities return to their initial values.

To locate these solutions, we define a Poincar\'e map that removes the accumulated rotations:
\begin{equation}
\begin{aligned}
    \mathbf{P}:\mathbf{X}_n&\mapsto\mathbf{X}_{n+1},\\
    \mathbf{X}_{n+1}&=\mathbf{F}(T,\mathbf{X}_n)
    -(2\pi w_x,0,2\pi w_y,0).
\end{aligned}
\end{equation}
A point on the rotatory orbit is therefore a fixed point of this map. Using the rotational periodicity condition~\eqref{eq:rb_period}, we solve the nonlinear system
\begin{equation}
    \mathbf{R}(T,\mathbf{X}_0)
    =\mathbf{F}(T,\mathbf{X}_0)-\mathbf{X}_0
    -(2\pi w_x,0,2\pi w_y,0)=\mathbf{0}.
    \label{eq:exist_syst_rb}
\end{equation}
The unknowns are the period $T$ and a point $\mathbf{X}_0^{(r)}$ on the desired rotatory trajectory. As for breathing orbits, the time-shift invariance makes one equation redundant. We remove this degeneracy by imposing a scalar phase condition, chosen here as $x(0)=0$.

The calculations below focus on the single-turn family with $w_x=w_y=-1$, for which both intercluster phase differences complete one full rotation during each period. Newton--Raphson iteration yields the period and the initial point of the rotatory orbit, after which numerical continuation in $\alpha_2$ and $\varepsilon_2$ traces the corresponding branch.

The full-network variational system~\eqref{eq:after_ABC} remains valid along a rotatory orbit because it was derived without assuming that $x(t)$ and $y(t)$ are bounded. We therefore integrate the Floquet system~\eqref{eq:floquet_system} over one period on the cylindrical phase space and calculate the eigenvalues of the resulting monodromy matrix. The rotobreathing state is stable when all nontrivial Floquet multipliers lie inside the unit circle. As for breathing orbits, one neutral multiplier is equal to unity because of time-translation invariance.

\subsection{Origin through global bifurcations and existence boundaries}
\label{sec:rotobreathing:birth}

\begin{figure}[t!]
	\centering
	\includegraphics[width=1.0\linewidth]{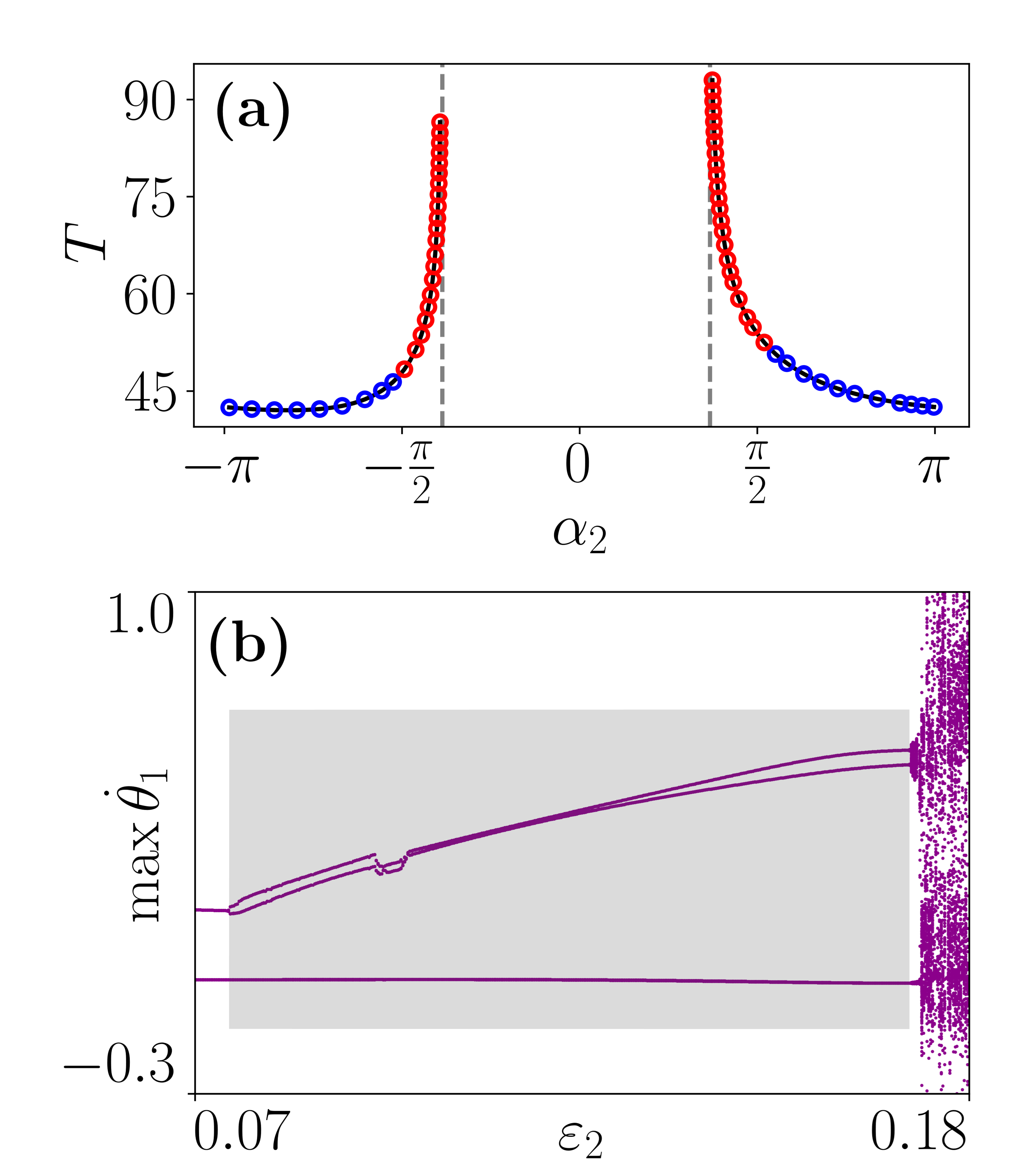}
	\caption{Bifurcation signatures of rotobreathing cyclops states: global merging with a heteroclinic-contour-like structure and period-doubling destabilization.
		(a) Period $T$ of the rotobreathing orbit as a function of $\alpha_2$ for fixed $\varepsilon_2=0.05$, corresponding to the dark-gray dotted line in Fig.~\ref{fig:br_stable_area}(b). Blue and red markers denote stable and unstable rotatory orbits, respectively. The period grows sharply near the dotted gray vertical asymptotes, which mark the boundaries of the rotobreathing existence intervals and support the interpretation of a global merging with a heteroclinic-contour-like saddle structure.
		(b) Local maxima of the instantaneous frequency $\dot{\theta}_1$ as a function of $\varepsilon_2$ for fixed $\alpha_2=2.65$, corresponding to the light-gray solid line in Fig.~\ref{fig:br_stable_area}(b). At the period-doubling point, the single sequence of maxima separates into two alternating branches. The gray rectangle marks the substantial parameter interval over which the stable period-doubled, phase-split rotobreathing state persists before the motion becomes irregular. Rotatory orbits are identified using the procedure described in Sec.~\ref{sec:rotobreathing:found}. Parameters: $N=11$, $\mu=1$, $\varepsilon_1=1$, and $\alpha_1=1.7$.}
	\label{fig:tree}
\end{figure}

Figure~\ref{fig:br_stable_area}(b) shows the continuation map for the same fixed parameters used for breathing states in Fig.~\ref{fig:br_stable_area}(a): $N=11$, $\mu=1$, $\varepsilon_1=1$, and $\alpha_1=1.7$. Blue regions indicate stable rotobreathing cyclops states, whereas red regions indicate rotatory orbits that exist but are unstable. The solid and dashed black curves mark secondary destabilization through Floquet-multiplier crossings at $\lambda_{\max}=-1$ and $\lambda_{\max}=+1$, respectively. The rotatory orbit continues as an unstable solution after crossing either black curve.

The gray curves with black crosses instead mark the existence boundaries of the rotobreathing branches. At these boundaries, the rotatory limit cycle merges with a rotatory heteroclinic-contour-like structure formed by two symmetry-related saddle two-cluster states. Traversed in one direction, this global bifurcation terminates the rotobreathing branch; traversed in the opposite direction, it gives birth to the rotatory orbit.

The phase-space geometry is analogous to that found for the global boundary of breathing states, but now the trajectory has nonzero winding numbers. As the boundary is approached, one of the intercluster phase differences remains close to zero for an increasingly long interval. When $x(t)\approx0$, the first coherent cluster nearly coincides with the solitary oscillator and the system approaches a $(K+1):K$ two-cluster configuration. When $y(t)\approx0$, it approaches the symmetry-related configuration $K:(K+1)$. These two-cluster states act as saddles for the reduced dynamics.

\begin{figure}[t!]
	\centering
	\includegraphics[width=1.0\linewidth]{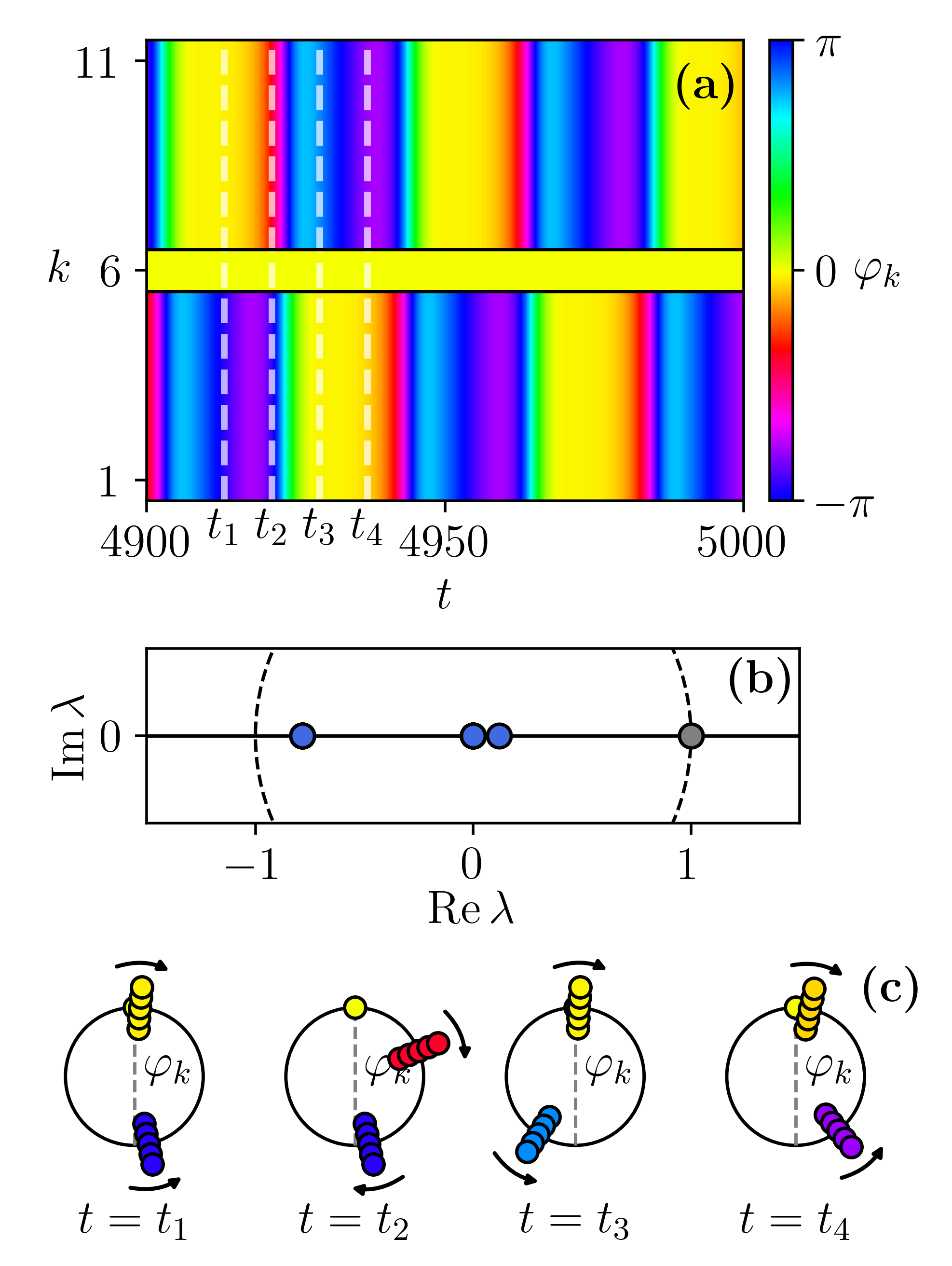}
	\caption{Stable rotobreathing cyclops state at point $D$ in Fig.~\ref{fig:br_stable_area}(b).
		(a) Phase differences $\varphi_k=\theta_k-\theta_6$ relative to the solitary oscillator. The two coherent clusters remain internally synchronized while completing periodic $2\pi$ rotations relative to the solitary oscillator. Vertical dashed lines mark the times used for the phase snapshots in panel (c).
		(b) Floquet multipliers of the rotatory orbit. All nontrivial multipliers, shown in blue, lie inside the unit circle; the gray multiplier at $\lambda=1$ is neutral.
		(c) Instantaneous phase configurations at $t_1=4913$, $t_2=4921$, $t_3=4929$, and $t_4=4937$. Arrows indicate the direction of rotation relative to the solitary oscillator. Parameters: $N=11$, $\mu=1$, $\varepsilon_1=1$, $\alpha_1=1.7$, $\varepsilon_2=0.08$, and $\alpha_2=-2.704$.}
	\label{fig:rb_stable_A_full}
\end{figure}

During each rotobreathing cycle, the trajectory spends a long time near one of the saddle configurations, then undergoes a relatively rapid $2\pi$ phase slip and approaches the other. As the global boundary is approached, the residence times near the two saddles grow, while the rapid rotational parts of the trajectory remain comparatively short. At the limiting parameter value, the rotatory cycle merges with a contour-like connection between the two saddle cluster states and ceases to exist.

\begin{figure*}[t!]
	\centering
	\includegraphics[width=0.70\linewidth]{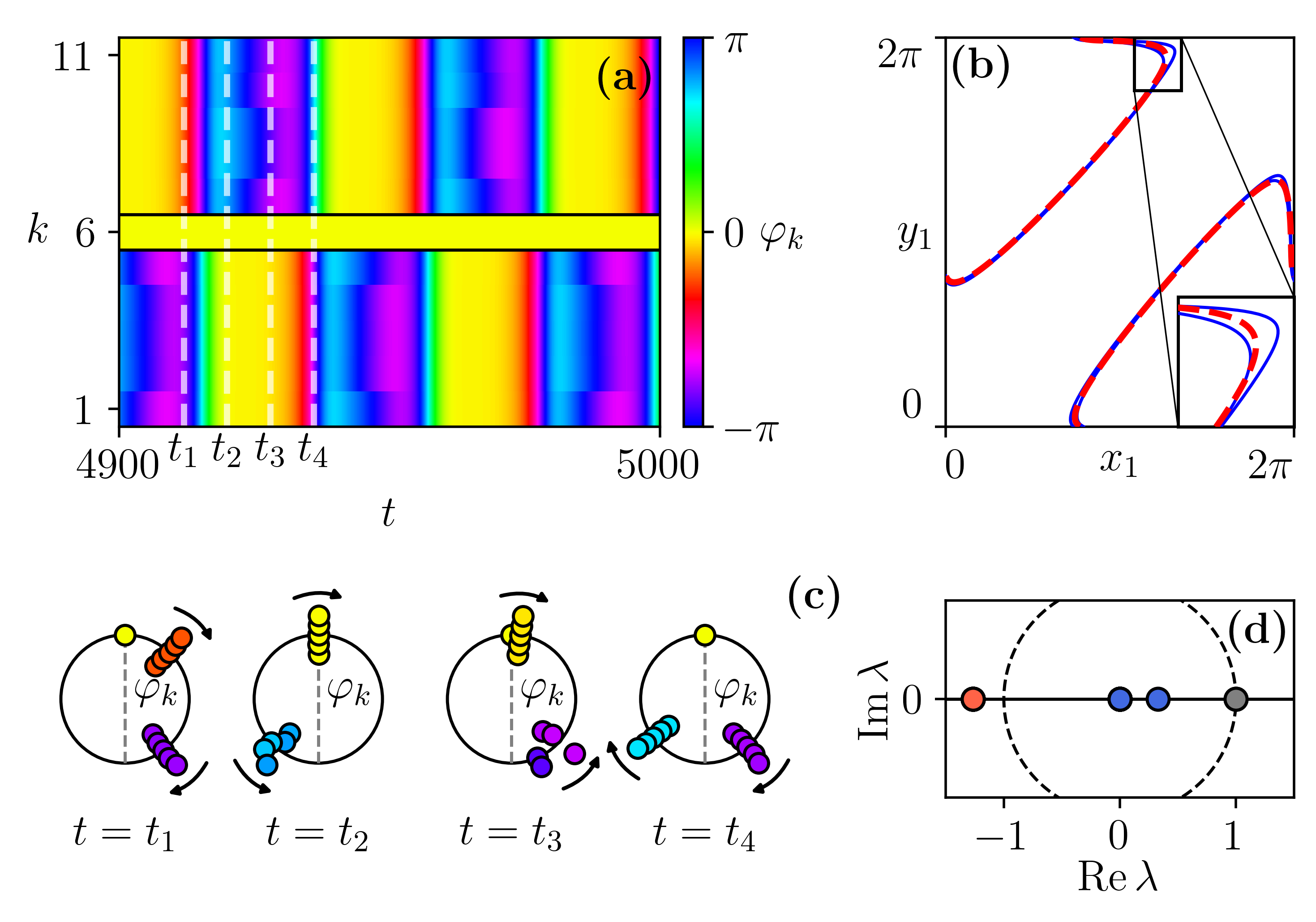}
	\caption{Period-doubling destabilization of a rotobreathing cyclops state and formation of a phase-split rotobreathing cyclops state at point $E$ in Fig.~\ref{fig:br_stable_area}(b).
		(a) Phase differences relative to the solitary oscillator. The parent clusters remain frequency locked, but a pronounced and persistent internal phase splitting develops in one cluster.
		(b) Projection in the reduced phase plane. The red dashed curve is the unstable period-$T$ rotatory orbit obtained by continuation, whereas the blue curve is the attracting period-$2T$ descendant. The inset resolves the double-loop structure.
		(c) Instantaneous phase configurations at $t_1=4912$, $t_2=4920$, $t_3=4928$, and $t_4=4936$. The internal phase split is especially clear at $t_2$ and $t_3$.
		(d) Floquet multipliers of the parent period-$T$ orbit. The red multiplier has crossed $-1$, the blue multipliers remain inside the unit circle, and the gray multiplier at $1$ is neutral. Parameters: $N=11$, $\mu=1$, $\varepsilon_1=1$, $\alpha_1=1.7$, $\varepsilon_2=0.08$, and $\alpha_2=2.65$.}
	\label{fig:rb_unstable_B_full}
\end{figure*}

\begin{figure*}[t!]
	\centering
	\includegraphics[width=0.70\linewidth]{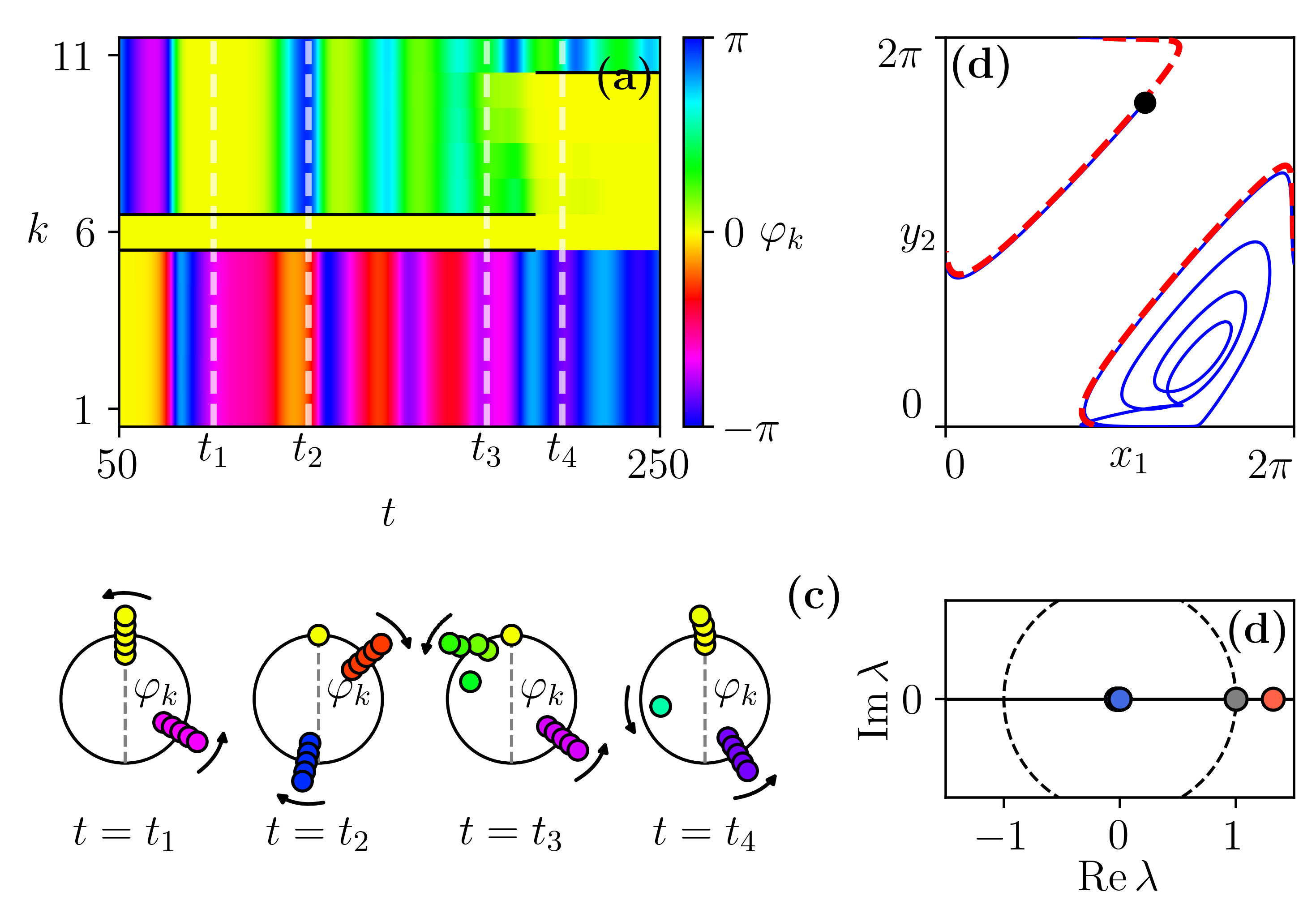}
	\caption{Cluster-destruction breakdown of a rotobreathing cyclops state and transition to switching dynamics at point $F$ in Fig.~\ref{fig:br_stable_area}(b).
		(a) Phase differences relative to the original solitary oscillator display the loss and reformation of coherent groups.
		(b) Projection in the reduced phase plane. The red dashed curve is the unstable rotobreathing orbit, the black dot marks the perturbed initial condition, and the blue trajectory departs from the orbit toward switching dynamics.
		(c) Instantaneous phase configurations at $t_1=85$, $t_2=120$, $t_3=186$, and $t_4=214$. The original solitary oscillator joins a coherent cluster, while another oscillator separates and assumes the solitary role.
		(d) Floquet multipliers of the parent orbit. The red multiplier has crossed $+1$, the blue multipliers remain inside the unit circle, and the gray multiplier at $1$ is neutral. Parameters: $N=11$, $\mu=1$, $\varepsilon_1=1$, $\alpha_1=1.7$, $\varepsilon_2=0.129$, and $\alpha_2=1.72$.}
	\label{fig:rb_unstable_C_full}
\end{figure*}

The growth of the rotation period provides direct numerical evidence for this interpretation. Figure~\ref{fig:tree}(a) shows the period $T$ as a function of $\alpha_2$ at fixed $\varepsilon_2=0.05$, along the dark-gray dotted line in Fig.~\ref{fig:br_stable_area}(b). The period increases sharply as either boundary of the rotobreathing existence interval is approached, with the dotted vertical lines indicating the corresponding asymptotic parameter values. This period growth is characteristic of a global bifurcation involving saddle connections.

Figure~\ref{fig:tree}(b) illustrates a complementary bifurcation signature of the rotobreathing branch, showing its period-doubling destabilization and the parameter interval occupied by the resulting phase-split rotobreathing state. This route is discussed in detail in Sec.~\ref{sec:rotobreathing:breakdown}.

Importantly, numerical continuation in Fig.~\ref{fig:br_stable_area}(b) generally reveals no connection between the rotobreathing branch and the Andronov--Hopf branch from which breathing states emerge. Rotobreathing cyclops states are therefore not the large-amplitude continuation of breathing cyclops states. Instead, they constitute a separate family of cyclops dynamics created through a global bifurcation. This mechanism parallels the global homoclinic bifurcation of a saddle equilibrium encountered in driven-pendulum descriptions of inertial intercluster and solitary-state dynamics.~\cite{belykh2016bistability,munyayev2022stability}

As a representative stable rotobreathing cyclops state, we consider point $D$ in Fig.~\ref{fig:br_stable_area}(b), with $\varepsilon_2=0.08$ and $\alpha_2=-2.704$. This point lies inside the blue stability region. The corresponding trajectory is a rotatory orbit in the cylindrical $(x,y)$ phase space, with both phase differences completing one full turn per period (Fig.~\ref{fig:phase_portraits}(f)). Direct numerical integration of the full system~\eqref{eq:model_def}, initialized near the identified rotatory orbit, produces the dynamics shown in Fig.~\ref{fig:rb_stable_A_full}.

The phase differences in Fig.~\ref{fig:rb_stable_A_full}(a) undergo repeated $2\pi$ rotations relative to the solitary oscillator, while the oscillators within each of the two coherent clusters remain perfectly phase synchronized. All nontrivial Floquet multipliers lie inside the unit circle (Fig.~\ref{fig:rb_stable_A_full}(b)), confirming the stability of the rotobreathing orbit. The snapshots in Fig.~\ref{fig:rb_stable_A_full}(c) illustrate the full rotations of the coherent clusters relative to the solitary oscillator.

\subsection{Destabilization routes of rotobreathing cyclops states}
\label{sec:rotobreathing:breakdown}

Despite their distinct global bifurcation origin, rotobreathing cyclops states lose stability through the same two generic real-multiplier scenarios as breathing cyclops states. A crossing at $\lambda_{\max}=-1$ produces a period-doubled, phase-split rotobreathing state, whereas a crossing at $\lambda_{\max}=+1$ destroys the coherence of a parent cluster and leads to switching or more general multicluster dynamics.

At point $E$ in Fig.~\ref{fig:br_stable_area}(b), with $\varepsilon_2=0.08$ and $\alpha_2=2.65$, the leading nontrivial Floquet multiplier crosses $-1$, and the parent period-$T$ rotatory orbit undergoes a period-doubling bifurcation. The resulting dynamics are shown in Fig.~\ref{fig:rb_unstable_B_full}. The phase-difference diagram in Fig.~\ref{fig:rb_unstable_B_full}(a) shows that the distinguished solitary oscillator persists and that the oscillators originating from the two parent clusters remain frequency locked. Exact phase synchrony within one cluster is nevertheless lost, producing persistent internal phase offsets and smaller phase-synchronized subclusters. We call the resulting regime a \emph{phase-split rotobreathing cyclops state}. In the reduced phase plane, the unstable parent orbit forms a single loop, whereas a small perturbation approaches a stable double-loop orbit with period $2T$ (Fig.~\ref{fig:rb_unstable_B_full}(b)). The complete intercluster phase configuration therefore no longer repeats after one period of the parent rotobreather; instead, it alternates over two rotational cycles before returning to the same configuration. In contrast to the weak phase splitting observed near the breathing-state period-doubling boundary, the splitting at point $E$ is pronounced and is clearly visible in the phase snapshots at $t_2$ and $t_3$ (Fig.~\ref{fig:rb_unstable_B_full}(c)). The Floquet spectrum in Fig.~\ref{fig:rb_unstable_B_full}(d) confirms that the parent orbit loses stability through a real multiplier crossing at $-1$.

The period-doubling route reveals a constructive role of self-generated phase disorder. When the perfectly phase-aligned rotobreathing cyclops state loses stability, a stable period-doubled branch emerges with small intracluster phase offsets. As this branch is followed away from the bifurcation, the phase splitting broadens while frequency locking, the distinguished solitary oscillator, and the overall cyclops organization remain intact. The resulting symmetry-broken phase distribution therefore stabilizes a descendant of the otherwise unstable parent cyclops state. This mechanism is reminiscent of the heterogeneity-induced stabilization of stationary cyclops states reported in Ref.~\cite{bolotov2025heterogeneity}, where imposed intrinsic-frequency disorder broadened the cluster phases and stabilized cyclops configurations that were unstable for identical oscillators. Here, however, no oscillator heterogeneity is introduced: the phase spread is generated endogenously by the collective dynamics through a period-doubling bifurcation. The effect can thus be viewed as phase-split-promoted stabilization of cyclops dynamics in a network of identical oscillators.

The local-maxima diagram in Fig.~\ref{fig:tree}(b) shows that the period-doubled state is not confined to a narrow neighborhood of the bifurcation. At fixed $\alpha_2=2.65$, the single sequence of local maxima of $\dot{\theta}_1$ splits into two alternating branches at the period-doubling point. The two branches persist over the substantial interval
$
0.075\lesssim\varepsilon_2\lesssim0.171,
$
demonstrating that the phase-split rotobreathing state occupies a sizable parameter range. Beyond this interval, the sequence of maxima becomes irregular, indicating a transition to more complex dynamics. Direct simulations in this region produce irregular multicluster behavior and suggest that the period-doubling route can provide an entry to potentially chaotic phase-offset dynamics. 

A second destabilization route occurs at point $F$ in Fig.~\ref{fig:br_stable_area}(b), with $\varepsilon_2=0.129$ and $\alpha_2=1.72$. Here, the leading Floquet multiplier crosses $+1$. The parent rotatory orbit continues as an unstable solution, but the associated perturbation destroys the coherence of one of its clusters (Fig.~\ref{fig:rb_unstable_C_full}(a-c)). In the reduced phase-plane projection in Fig.~\ref{fig:rb_unstable_C_full}(b), the red dashed curve represents the unstable rotobreathing orbit, while the blue trajectory initiated near it departs toward switching dynamics.

The corresponding full-network dynamics are shown in Figs.~\ref{fig:rb_unstable_C_full}(a) and~\ref{fig:rb_unstable_C_full}(c). The original solitary oscillator approaches and joins one of the coherent clusters. An oscillator from that cluster then separates and becomes the new solitary unit. This process repeats, producing recurrent reorganization of the cyclops configuration and a switching cyclops state. Depending on the parameters, the same cluster-destruction route may lead instead to more general multicluster dynamics.

Unlike period doubling, the $+1$ instability represents a genuine breakdown of the parent rotobreathing cyclops state because its original coherent-cluster composition is destroyed. The Floquet spectrum in Fig.~\ref{fig:rb_unstable_C_full}(d) confirms that this breakdown is initiated by a real multiplier crossing through $+1$.

The two destabilization routes therefore reproduce the generic scenarios found for breathing states despite the distinct global origin of rotobreathers. A crossing at $-1$ preserves the overall cyclops organization but doubles the period and produces a stable phase-split rotobreathing state. A crossing at $+1$ destroys a coherent cluster and causes the parent rotobreather to break down into switching or multicluster dynamics.

The broad stability regions in Fig.~\ref{fig:br_stable_area}(b), together with the sizable period-doubled interval in Fig.~\ref{fig:tree}(b), show that rotobreathing cyclops states and their descendants form an important part of the network dynamics rather than isolated bifurcation branches. In particular, many of these regions lie in the overall repulsive-coupling domain, where complete synchronization is unstable. In the following section, we examine their attraction basins and show that breathing and rotobreathing cyclops states can become prevalent and, in some parameter intervals, effectively global attractors.

\section{Prevalence and persistence across network sizes}\label{sec:persistence}

\begin{figure}[t!]
    \centering
    \includegraphics[width=1.0\linewidth]{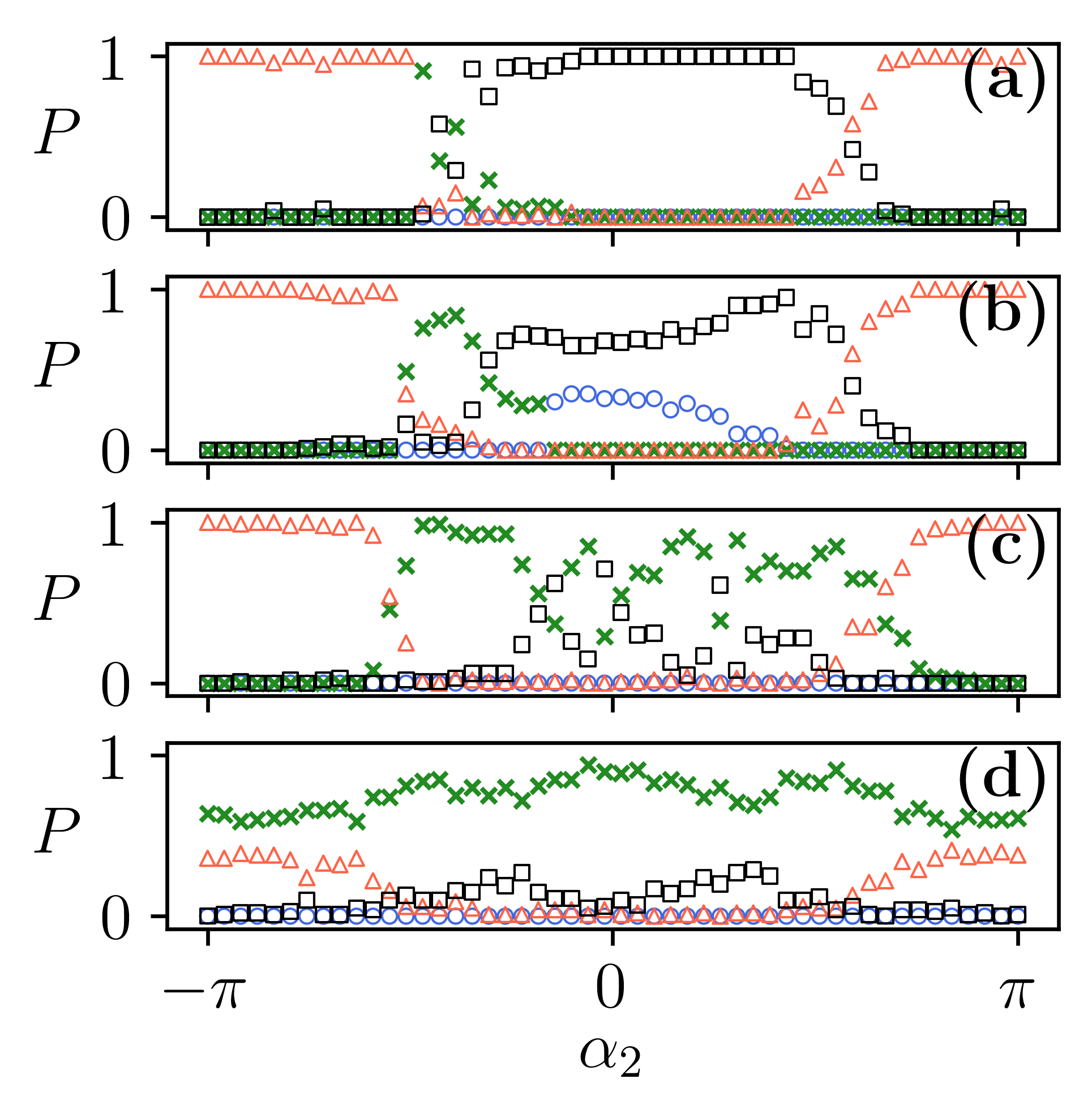}
 \caption{Estimated attraction-basin fractions $P$ as functions of the second-harmonic phase lag $\alpha_2$. Blue circles indicate stationary cyclops states, green crosses breathing cyclops states, and red triangles rotobreathing cyclops states. Black squares denote all other outcomes, including switching and multicluster states. Rotobreathing cyclops states dominate broad parameter intervals for the moderate and strong second-harmonic couplings in panels (a)--(c), whereas breathing cyclops states become prevalent for the weak second-harmonic coupling in panel (d). Each point is based on $100$ random initial conditions with $\theta_k(0)\in[-\pi,\pi]$ and $\dot{\theta}_k(0)\in[-1,1]$. Parameters: $N=11$, $\mu=1$, $\varepsilon_1=1$, and $\alpha_1=1.7$; (a) $\varepsilon_2=0.15$, (b) $\varepsilon_2=0.10$, (c) $\varepsilon_2=0.05$, and (d) $\varepsilon_2=0.01$.}
    \label{fig:br_probability}
\end{figure}

The continuation diagrams establish local stability but do not indicate how readily the corresponding states are reached from generic initial conditions. We therefore estimate their attraction-basin fractions by integrating Eq.~\eqref{eq:model_def} from $100$ independent random initial conditions at each parameter value. The initial phases are sampled uniformly from $[-\pi,\pi]$, and the initial angular velocities from $[-1,1]$. Figure~\ref{fig:br_probability} shows the resulting probabilities as functions of $\alpha_2$ for four fixed values of $\varepsilon_2$. Breathing and rotobreathing cyclops states dominate over substantial parameter intervals, demonstrating that the periodic and rotatory branches are not isolated continuation artifacts but prevalent attractors of the full network. In particular, for the parameter ranges shown in Fig.~\ref{fig:br_probability}(a-c), rotobreathing states are realized with probabilities approaching unity within the overall repulsive-coupling regime, where complete phase synchronization is unstable. Only for very weak second-harmonic coupling do breathing cyclops states instead become dominant (Fig.~\ref{fig:br_probability}(d)).

The odd network size is essential to this organization. For $N=2K+1$, a solitary oscillator separates two equal coherent clusters and supports the $K:1:K$ cyclops solution. In even-sized networks, the dominant alternatives are typically a stationary symmetric antiphase two-cluster state and, where stable, complete synchronization; their exact stability conditions are derived in Appendix~\ref{sec:appendix_even_N}. Even-sized networks may also admit asymmetric cyclops states of the form $(K-1):1:K$, with coherent clusters whose sizes differ by one. As shown in Appendix~\ref{sec:cyclops_even_N}, however, these states occupy only a very narrow stability region and have a small attraction basin. Consequently, adding or removing a single oscillator can replace prevalent breathing and rotobreathing dynamics with predominantly stationary collective states.

\begin{figure}[t!]
    \centering
    \includegraphics[width=1.0\linewidth]{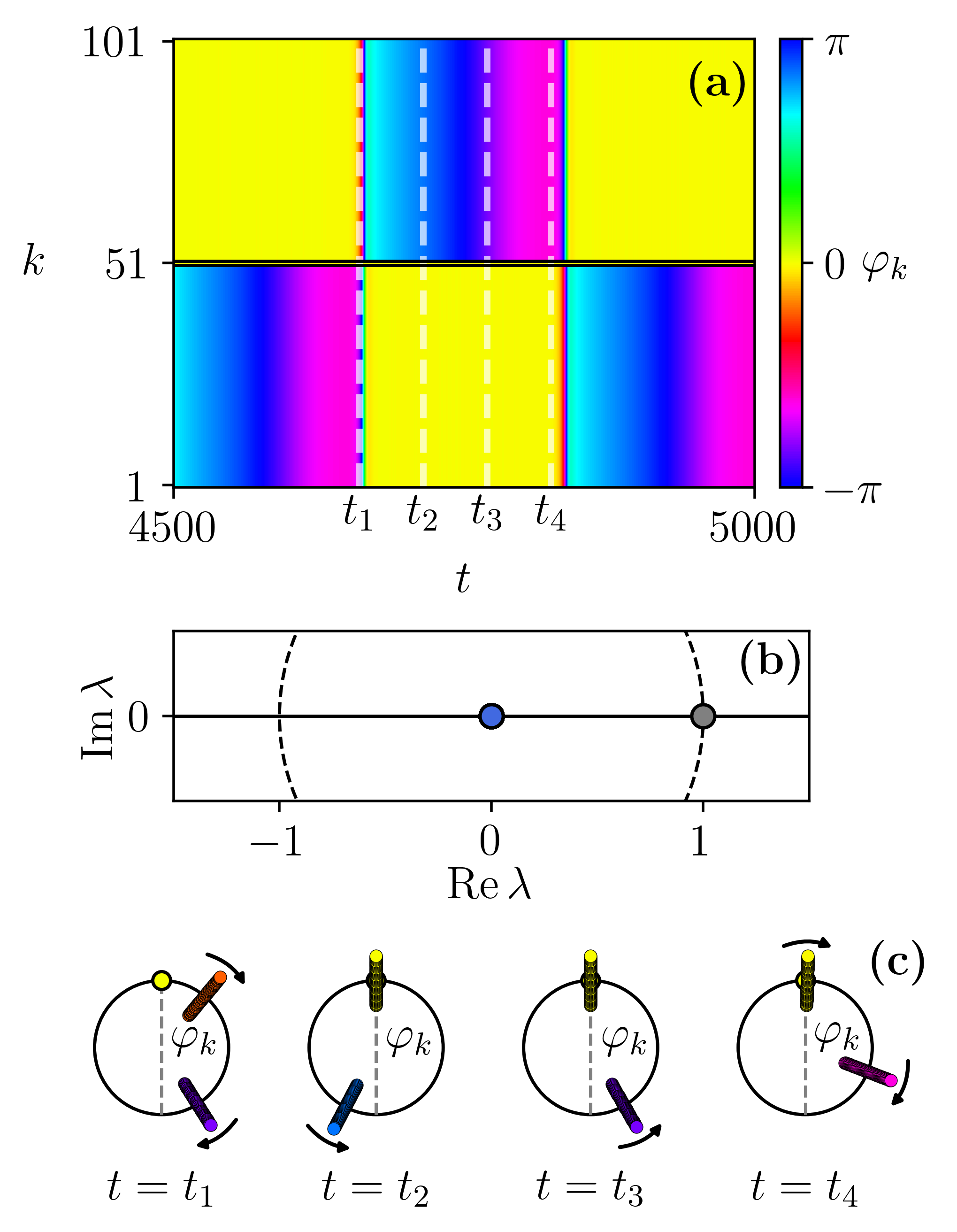}
    \caption{Stable rotobreathing cyclops state in a large odd-sized network. (a) Phase differences $\varphi_k=\theta_k-\theta_{51}$ for $N=101$; the two clusters remain coherent and alternate between prolonged near-locking to the solitary oscillator and rapid phase slips. (b) Floquet multipliers, with all nontrivial multipliers inside the unit circle. (c) Instantaneous phase configurations at $t_1=4660$, $t_2=4715$, $t_3=4770$, and $t_4=4825$. Parameters: $\mu=1$, $\varepsilon_1=1$, $\alpha_1=1.6$, $\varepsilon_2=0.06$, and $\alpha_2=2.0$.}
    \label{fig:rb_largeN}
\end{figure}

\begin{figure}[t!]
    \centering
    \includegraphics[width=1.0\linewidth]{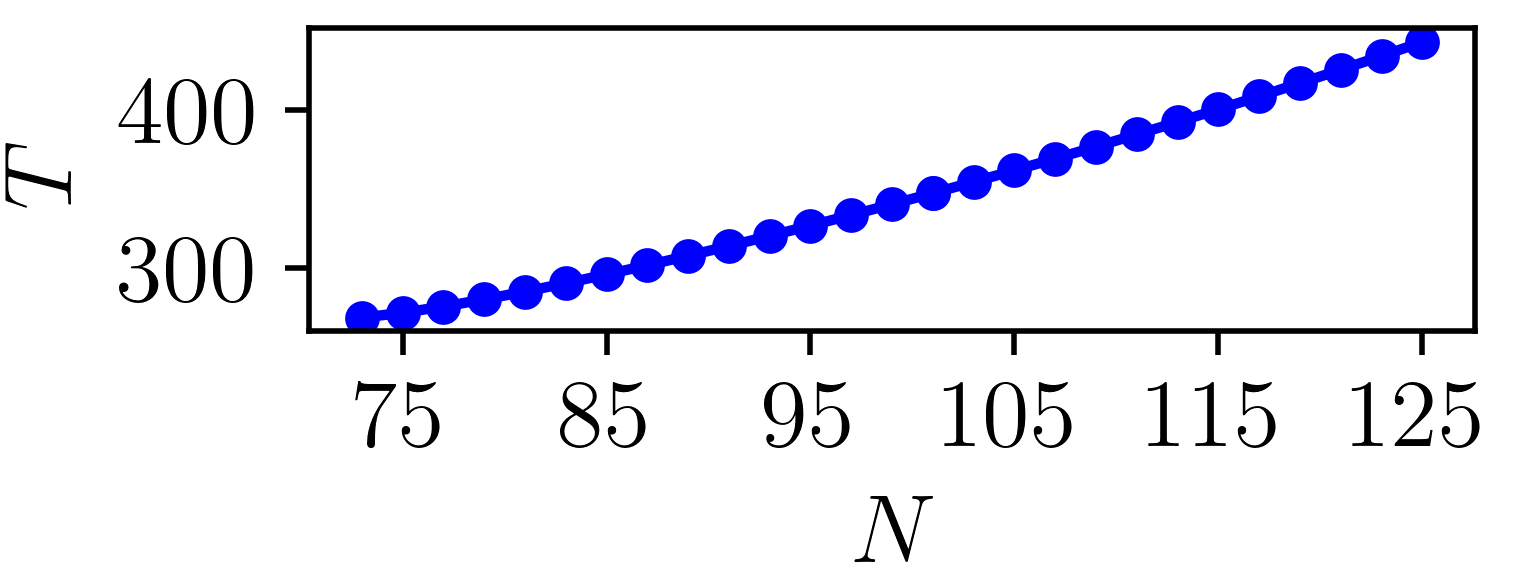}
    \caption{Period $T$ of the stable rotobreathing cyclops state as a function of odd network size $N$. The period increases with $N$ because the clusters spend progressively longer intervals near the solitary oscillator before completing a rapid phase slip. Parameters held fixed are $\mu=1$, $\varepsilon_1=1$, $\alpha_1=1.6$, $\varepsilon_2=0.06$, and $\alpha_2=2.0$.}
    \label{fig:rb_T_by_N}
\end{figure}

Cyclops motion with time-dependent intercluster phases also persists far beyond the small network used for continuation. Figure~\ref{fig:rb_largeN} shows a stable rotobreathing state for $N=101$. 
During each cycle, one coherent cluster remains nearly phase locked to the solitary oscillator for an extended interval and then undergoes a rapid $2\pi$ phase slip, after which the other cluster assumes the near-locked position. Thus, the trajectory alternately approaches the two symmetry-related $(K+1):K$ and $K:(K+1)$ two-cluster configurations before completing each rotation. These long residence intervals become increasingly pronounced with network size. Correspondingly, Fig.~\ref{fig:rb_T_by_N} shows that the period of the stable rotobreathing state grows as $N$ increases. Since an increasing period is a characteristic signature of approaching the global existence boundary, this trend suggests that increasing the network size moves the rotobreathing state closer to the heteroclinic contour-like bifurcation. The prolonged near-locking and rapid phase slips visible in Fig.~\ref{fig:rb_largeN} provide additional support for the role of the saddle two-cluster configurations and their heteroclinic contour-like connection in organizing rotobreathing dynamics. At the same time, the solitary oscillator remains dynamically active even in large odd-sized networks rather than being absorbed permanently into one of the coherent clusters.

\section{Conclusion}\label{sec:conclusion}

We developed a systematic bifurcation framework for the emergence, transformation, and breakdown of breathing and rotobreathing cyclops states in globally coupled networks of identical Kuramoto--Sakaguchi rotators with inertia and two-harmonic coupling. Stationary cyclops states are equilibria on a symmetry-defined $K:1:K$ synchronization manifold, whose existence, like that of other cluster states in identical oscillator networks, is determined by network symmetries and balanced input structure.~\cite{belykh2000hierarchy,pogromsky2002partial,golubitsky2006nonlinear,belykh2011mesoscale} Their loss of stability replaces the fixed intercluster geometry with time-dependent collective motion. Characterizing the resulting states is substantially more challenging because breathing cyclops states are periodic orbits with an unknown period, whereas rotobreathers are rotatory periodic orbits with nonzero winding numbers on a cylindrical phase space. By combining a reduction to the invariant $K:1:K$ manifold with Newton--Raphson searches for periodic and rotatory orbits, continuation in the second-harmonic parameters, and Floquet analysis of the full oscillator network, we followed both stable and unstable branches and distinguished bifurcations of the intercluster motion from instabilities that split or destroy the coherent clusters.

The two families have predominantly distinct origins but share common secondary instabilities. Breathing cyclops states are typically born locally from stationary cyclops states through an Andronov--Hopf bifurcation. A less common route involves a global bifurcation in which a breathing orbit emerges from a heteroclinic-contour-like structure formed by saddle cluster states. Rotobreathing states, by contrast, are organized primarily by such global saddle-connection mechanisms. Numerical continuation generally reveals rotobreathing branches that are separate from the Andronov--Hopf branches of breathing states, although isolated connections between the two families may occur. Thus, rotobreathers are not, in general, simply large-amplitude continuations of breathers. The rapid growth of their rotation period near the global bifurcation existence boundary further supports the heteroclinic-contour interpretation. Once formed, both breathing and rotobreathing states lose stability through the same two real-multiplier scenarios. A crossing at $-1$ destabilizes the parent orbit while preserving the overall cyclops organization, producing a stable period-doubled, phase-split descendant. A crossing at $+1$ destroys the coherence of a parent cluster and causes a genuine breakdown into switching cyclops or more general multicluster dynamics.

These bifurcation routes also help answer which rhythms prevail when the overall coupling is repulsive and complete synchronization is unstable. Attraction-basin statistics show that breathing and, especially, rotobreathing cyclops states dominate broad parameter intervals and can attain probabilities close to unity in the sampled initial-condition ensembles. They persist in large odd-sized networks, where the solitary oscillator remains dynamically active rather than being absorbed into a coherent cluster. Even-sized networks instead favor stationary antiphase two-cluster or fully synchronous states. Cyclops states and their time-dependent descendants may therefore provide a structural foundation for organizing repulsive Kuramoto dynamics, analogous to the organizing role of complete synchronization in attractively coupled networks. The mechanism relies on a division of roles: inertia enables oscillatory and rotatory intercluster motion, higher coupling harmonics stabilize the cyclops partition, and the solitary oscillator mediates the relative motion of the coherent groups.

The period-doubling route reveals an additional constructive role of symmetry breaking. As the period-doubled branch is followed away from its onset, dynamically generated intracluster phase offsets broaden while frequency locking, the solitary oscillator, and the coarse cyclops structure remain intact. The resulting phase-split distribution stabilizes a descendant of an otherwise unstable phase-aligned parent state. This effect complements the heterogeneity-induced stabilization of stationary cyclops states:~\cite{bolotov2025heterogeneity} there, externally imposed intrinsic-frequency disorder broadens the cluster phases and creates stability, whereas here the phase spread is generated endogenously by collective dynamics in a network of identical oscillators. The effect can therefore be viewed as phase-split-promoted stabilization of cyclops dynamics. It connects the present results to the broader literature on disorder-promoted collective dynamics and converse symmetry breaking,~\cite{nishikawa2016symmetric,zhang2021random,molnar2021asymmetry} including disorder-assisted coherence in laser arrays.~\cite{nair2021using,spitz2026coherent}

Several analytical and dynamical questions remain open. The reduced phase-difference equations have the structure of two coupled driven-pendulum-like systems and may support mixed-mode and chaotic intercluster motion,\cite{brister2020three} including Shilnikov-type spiral chaos generated by a homoclinic orbit of a saddle focus.~\cite{shilnikov1965case} Establishing the exact saddle connections underlying the heteroclinic-contour-like boundaries and deriving parameter-explicit conditions for their existence are natural subjects for future work. Auxiliary-system methods developed for inertial Kuramoto networks provide a possible analytical route.~\cite{brister2020three,barabash2021partial}

Finally, a recently developed predictive reduction maps time-delayed Kuramoto--Daido networks to delay-free networks of inertial rotators with delay-induced triadic interactions.~\cite{smirnov2026delay} Because this reduction makes time delay appear through effective inertia and higher-order interactions, the emergence, phase-split stabilization, and cluster-destruction mechanisms identified here may have counterparts in time-delayed oscillator networks. Testing this correspondence for breathing and rotobreathing cyclops states and determining how delay reshapes their local and global bifurcations are promising directions for future study.

\begin{acknowledgments}
This work was supported by the RSF under Project No.~22-12-00348-P (L.A.S., Sections II and III) and Project No.~24-72-00105 (M.M.K. and M.I.B., Sections IV and V), and the National Science Foundation (USA) under Grant DMS-2510860 (I.B.).
\end{acknowledgments}

\section*{Data Availability}
The numerical data and custom simulation codes that support the findings of this paper are available from the corresponding author upon reasonable request.

\appendix

\section{Symmetric two-cluster and synchronous states in even-sized networks}\label{sec:appendix_even_N}

We consider Eq.~\eqref{eq:model_def} for an even number of oscillators, $N=2K$. A principal collective regime in this case is the symmetric antiphase two-cluster configuration, in which two clusters $C_1$ and $C_2$ of size $K$ are internally synchronized and separated by $\pi$:
\begin{equation}
    D(2)=
    \begin{cases}
        \theta_k(t)=\Omega t+\pi, & k\in C_1,\\
        \theta_k(t)=\Omega t, & k\in C_2.
    \end{cases}
    \label{eq:2_clast_st}
\end{equation}
The common rotation frequency follows from self-consistency,
\[
    \Omega=\omega+\frac{1}{2}\big[H(\pi)+H(0)\big],
\]
where
$H(\pi)=H(-\pi)=\varepsilon_1\sin\alpha_1-\varepsilon_2\sin\alpha_2$
and
$H(0)=-\varepsilon_1\sin\alpha_1-\varepsilon_2\sin\alpha_2$.
In the frame rotating with $\Omega$, the state is the equilibrium
$\varphi_k^0=\pi$ for $k\in C_1$ and $\varphi_k^0=0$ for $k\in C_2$.

With $\varphi_k=\varphi_k^0+\delta\varphi_k$, linearization gives
\[
    \mu\,\delta\ddot{\varphi}_k+\delta\dot{\varphi}_k
    =\frac{1}{N}\sum_{n=1}^{N}
    H'(\varphi_n^0-\varphi_k^0)
    (\delta\varphi_n-\delta\varphi_k).
\]
Introduce
\begin{equation}
\begin{aligned}
    a&=H'(0)=\varepsilon_1\cos\alpha_1+2\varepsilon_2\cos\alpha_2,\\
    b&=H'(\pi)=H'(-\pi)=-\varepsilon_1\cos\alpha_1+2\varepsilon_2\cos\alpha_2.
\end{aligned}
\label{eq:ab_even}
\end{equation}
Pairs within the same cluster contribute $a$, while pairs in different clusters contribute $b$. In vector form,
\[
    \mu\,\delta\ddot{\boldsymbol{\varphi}}
    +\delta\dot{\boldsymbol{\varphi}}
    +\frac{1}{N}L\,\delta\boldsymbol{\varphi}=0,
\]
where
\[
(L\,\delta\boldsymbol{\varphi})_k
=K(a+b)\delta\varphi_k
-a\sum_{n\in C(k)}\delta\varphi_n
-b\sum_{n\notin C(k)}\delta\varphi_n,
\]
and $C(k)$ denotes the cluster containing oscillator $k$.

The spectrum follows from symmetry. A uniform phase shift gives the neutral eigenvalue $\tilde\lambda_0=0$. The intercluster perturbation, equal to $+1$ on $C_1$ and $-1$ on $C_2$, has eigenvalue $\tilde\lambda_1=2Kb$. Perturbations with zero mean in each cluster form an $(N-2)$-dimensional subspace with eigenvalue $\tilde\lambda_2=K(a+b)$. After division by $N=2K$, the nontrivial modal coefficients are
\begin{equation}
\begin{aligned}
    \lambda_1&=b &&\text{(multiplicity 1)},\\
    \lambda_2&=\frac{a+b}{2} &&\text{(multiplicity $N-2$)}.
\end{aligned}
\label{eq:even_eigenvalues}
\end{equation}
Each mode satisfies $\mu\ddot{z}+\dot{z}+\lambda z=0$ and is asymptotically stable if and only if $\lambda>0$. The symmetric antiphase state is therefore stable precisely when
\begin{equation}
    2\varepsilon_2\cos\alpha_2>\varepsilon_1\cos\alpha_1,
    \qquad
    \varepsilon_2\cos\alpha_2>0.
    \label{eq:stab_anti}
\end{equation}
The first inequality preserves the antiphase separation of the cluster centers; the second preserves coherence within each cluster. The thresholds do not depend on $N$ or $\mu$. Inertia changes the transient decay from monotone to oscillatory and modifies the relaxation time, but not the stability boundaries.

For comparison, the fully synchronous state has frequency
$\Omega_s=\omega-\varepsilon_1\sin\alpha_1-\varepsilon_2\sin\alpha_2$.
Its nonuniform perturbations all have modal coefficient $a$, so complete synchronization is stable if and only if
\begin{equation}
    a>0
    \quad\Longleftrightarrow\quad
    \varepsilon_1\cos\alpha_1+2\varepsilon_2\cos\alpha_2>0.
    \label{eq:stab_sync}
\end{equation}
Equations~\eqref{eq:stab_anti} and \eqref{eq:stab_sync} allow the symmetric two-cluster and synchronous states to coexist stably.

\begin{figure*}[t!]
    \centering
    \includegraphics[width=0.85\linewidth]{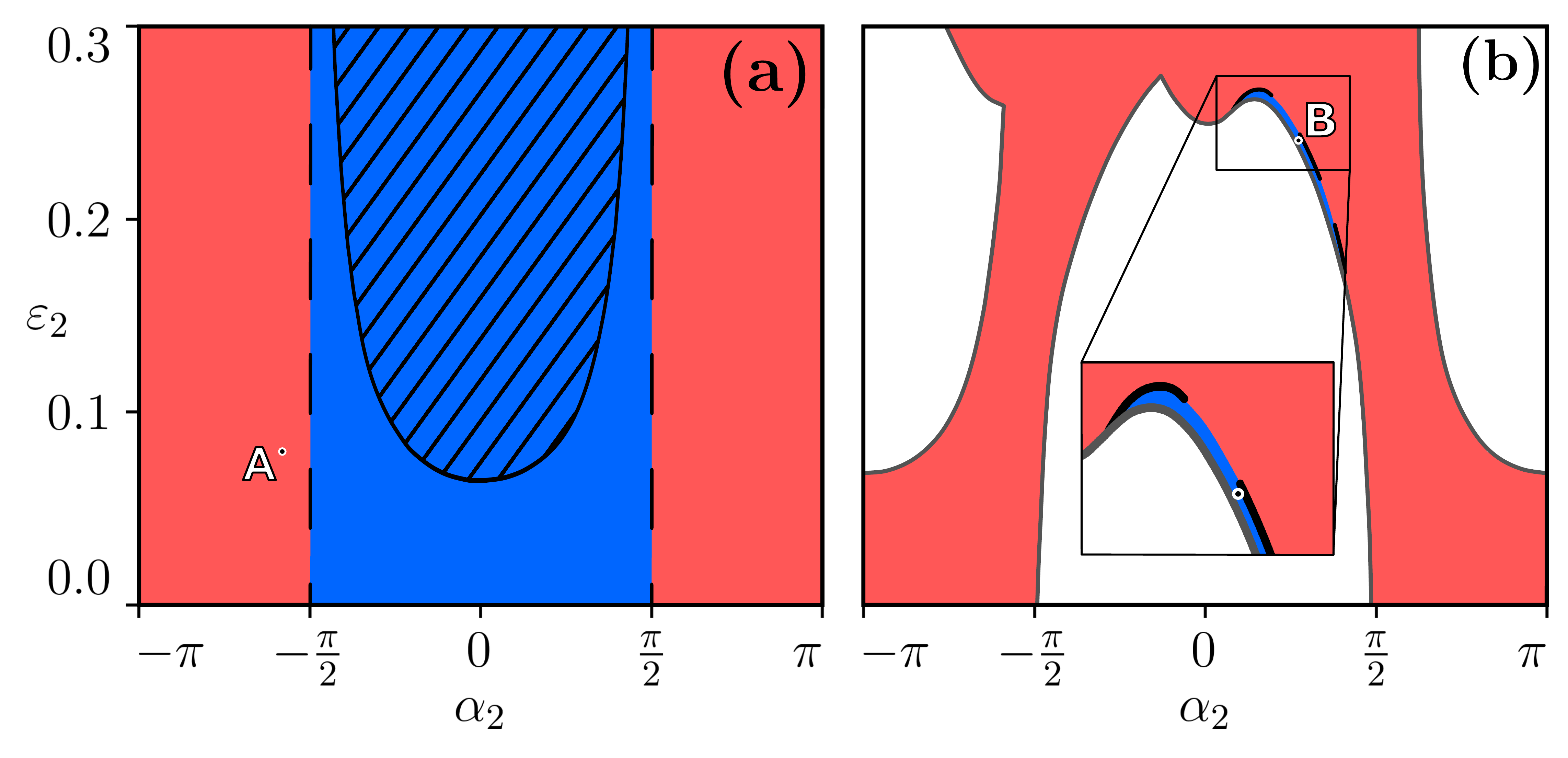}
     \caption{Stationary cluster states in even-sized networks.
    (a) Stability domains of the symmetric antiphase two-cluster state $N/2:N/2$ and the fully synchronous one-cluster state. Blue and red indicate stable and unstable symmetric two-cluster states, respectively. The dashed black line marks the stability boundary given by Eq.~\eqref{eq:stab_anti}, while the hatched region indicates stable complete synchronization according to Eq.~\eqref{eq:stab_sync}. Their overlap is the bistability region.
    (b) Existence and stability domains of stationary asymmetric $4:1:5$ cyclops states for $N=10$. Blue and red indicate stable and unstable states, respectively. The asymmetric cyclops states are obtained numerically from the reduced even-$N$ equations introduced in Appendix~\ref{sec:cyclops_even_N}, and their stability is evaluated in the full oscillator network. Points $A$ and $B$ correspond to the examples in Fig.~\ref{fig:tc_destroyed} and     Fig.~\ref{fig:asym_cyclops} in Appendix~\ref{sec:cyclops_even_N}, respectively. Parameters: $\mu=1$, $\varepsilon_1=1$, and $\alpha_1=1.7$.}
    \label{fig:even_osc}
\end{figure*}

\begin{figure}[t!]
    \centering
    \includegraphics[width=1.0\linewidth]{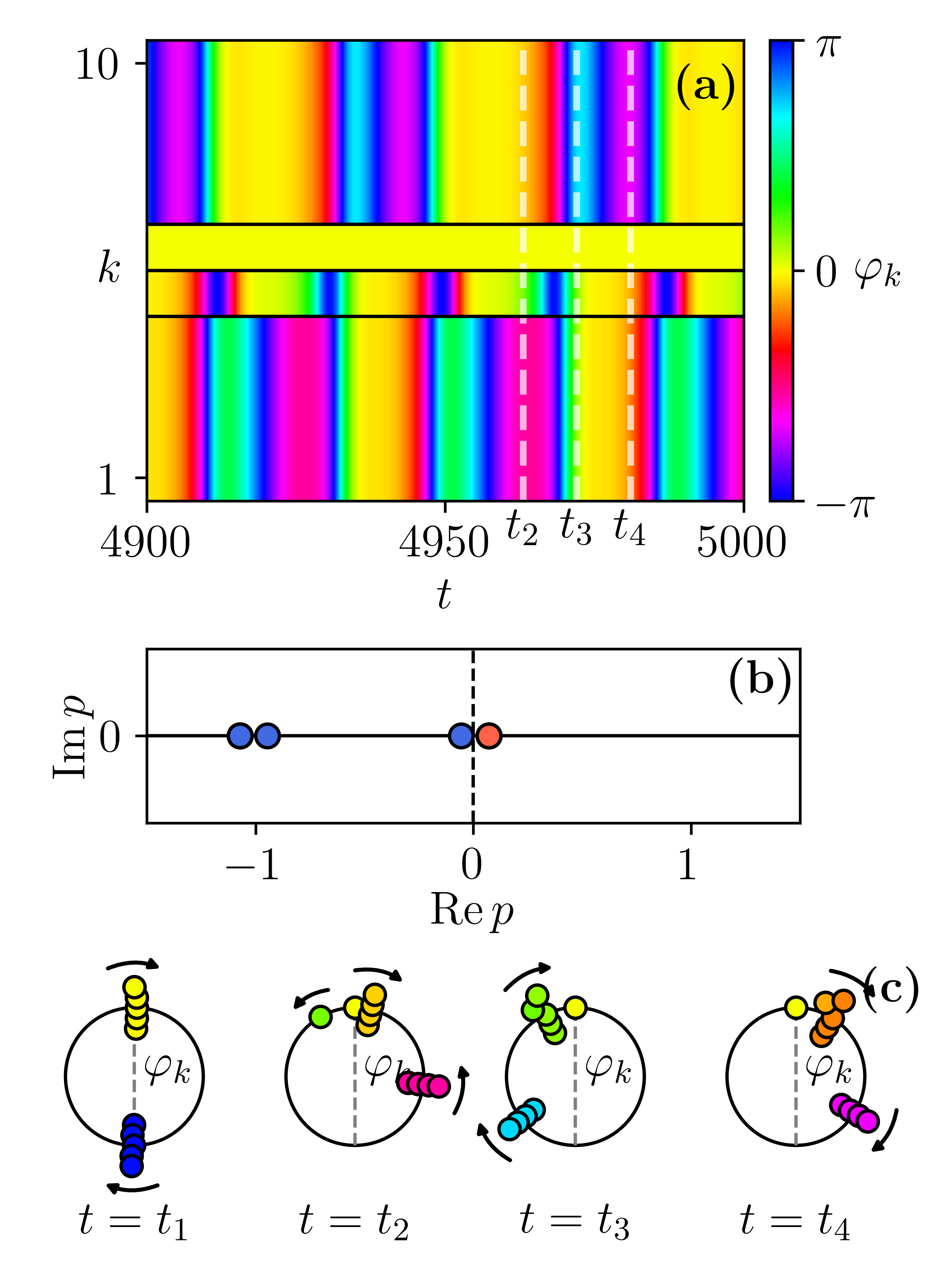}
    \caption{Loss of the symmetric antiphase two-cluster state and formation of a four-cluster state at point $A$ of Fig.~\ref{fig:even_osc}(a). (a) Phase differences $\varphi_k=\theta_k-\theta_1$ show both parent clusters splitting. (b) Characteristic exponents $p_i$ of the parent stationary state; blue points are stable and red points are unstable. (c) Instantaneous phase configurations at $t_1=10$, $t_2=4963$, $t_3=4972$, and $t_4=4981$. Parameters: $N=10$, $\mu=1$, $\varepsilon_1=1$, $\alpha_1=1.7$, $\varepsilon_2=0.08$, and $\alpha_2=-2.071$.}
    \label{fig:tc_destroyed}
\end{figure}

For the parameters used in the main text, $\alpha_1=1.7$ and $\varepsilon_1=1$, the antiphase state is stable for $\alpha_2\in(-\pi/2,\pi/2)$. Complete synchronization becomes stable above the curve $\varepsilon_2=0.0645/\cos\alpha_2$, producing the bistable hatched region in Fig.~\ref{fig:even_osc}(a). When the intracluster condition in Eq.~\eqref{eq:stab_anti} is violated, both parent clusters lose coherence; the example at point $A$ evolves to a four-cluster state (Fig.~\ref{fig:tc_destroyed}).

\section{Asymmetric cyclops-like states in even-sized networks}\label{sec:cyclops_even_N}

\begin{figure}[t!]
    \centering
    \includegraphics[width=0.90\linewidth]{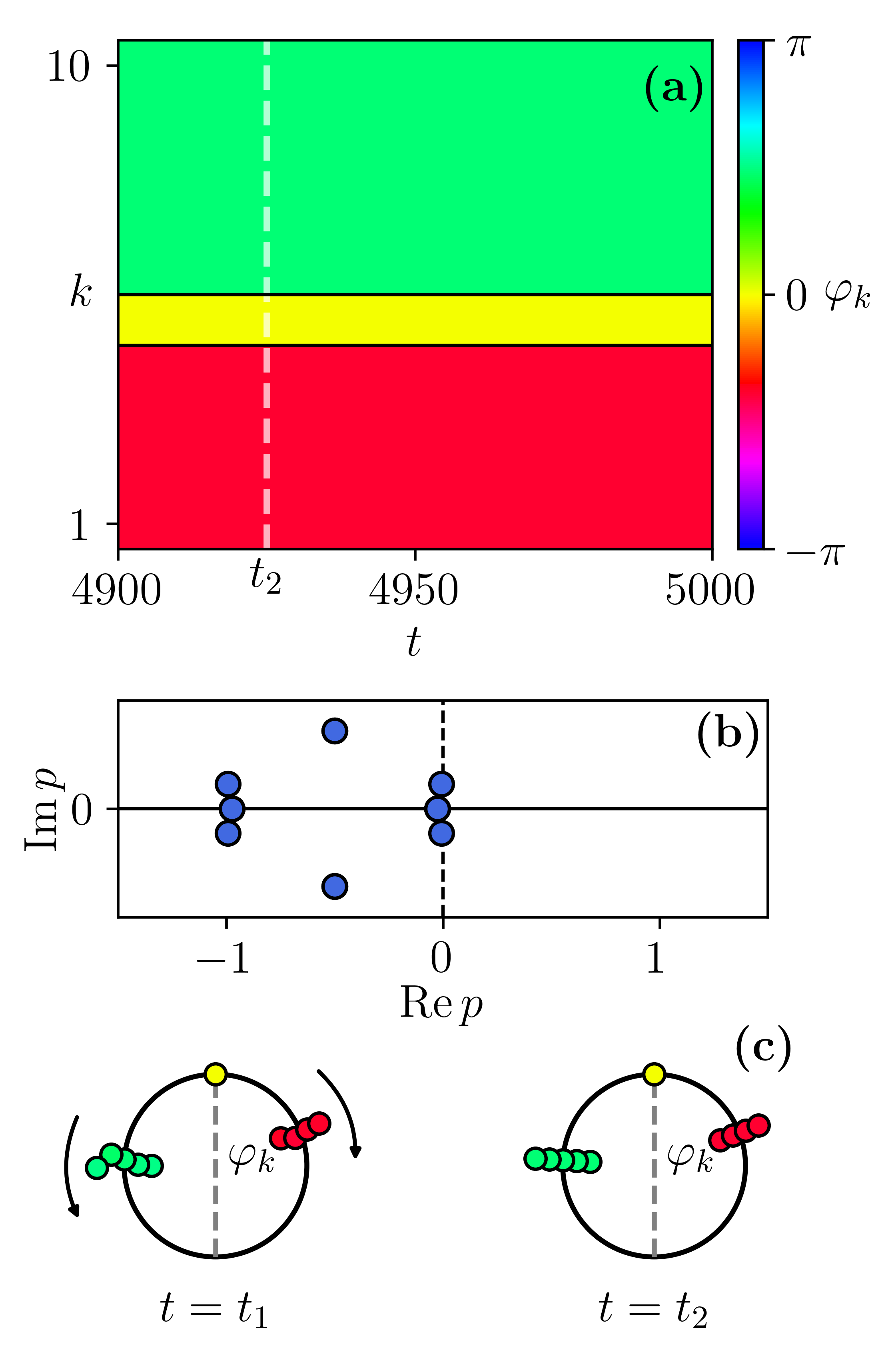}
    \caption{Stable asymmetric cyclops-like state with cluster partition $4:1:5$ at point $B$ of Fig.~\ref{fig:even_osc}(b). (a) Phase differences $\varphi_k=\theta_k-\theta_6$ relative to the solitary oscillator show two coherent clusters of unequal size. (b) Characteristic exponents $p_i$ of the stationary state; all nontrivial exponents have negative real parts. (c) Instantaneous phase configurations at $t_1=0.1$ and $t_2=4925$. Parameters: $N=10$, $\mu=1$, $\varepsilon_1=1$, $\alpha_1=1.7$, $\varepsilon_2=0.24$, and $\alpha_2=0.9$.}
    \label{fig:asym_cyclops}
\end{figure}

An even-sized network, $N=2K$, can also support the asymmetric three-cluster partition $(K-1):1:K$. It contains one solitary oscillator and two coherent clusters whose sizes differ by one. Let $x$ and $y$ denote the phase differences of the two clusters relative to the solitary oscillator. Subtracting the solitary-oscillator equation from the two cluster equations yields
\begin{widetext}
\begin{equation}
\begin{aligned}
    \mu\ddot{x}+\dot{x}
    &=\frac{1}{N}\sum_{q=1}^{2}\varepsilon_q\Big[
    -\sin(qx+\alpha_q)-(K-1)\sin(qx-\alpha_q)
    +K\sin(q(y-x)-\alpha_q)\\
    &\hspace{3.0cm}-K\sin(qy-\alpha_q)-(K-2)\sin\alpha_q\Big],\\
    \mu\ddot{y}+\dot{y}
    &=\frac{1}{N}\sum_{q=1}^{2}\varepsilon_q\Big[
    -\sin(qy+\alpha_q)-K\sin(qy-\alpha_q)\\
    &\hspace{3.0cm}+(K-1)\big(\sin(q(x-y)-\alpha_q)
    -\sin(qx-\alpha_q)-\sin\alpha_q\big)\Big].
\end{aligned}
\label{eq:xy_dyn_even}
\end{equation}
\end{widetext}
Stationary asymmetric cyclops-like states are equilibria of Eq.~\eqref{eq:xy_dyn_even}. We continue these equilibria in $(\alpha_2,\varepsilon_2)$ and evaluate their stability from the linearization of the full oscillator network. For $N=10$, the stable $4:1:5$ region is narrow (Fig.~\ref{fig:even_osc}(b)). Direct simulations also indicate a small local basin: the state in Fig.~\ref{fig:asym_cyclops} is reached only when perturbations from the equilibrium are of order $10^{-2}$ for the sampling used here. These asymmetric states are therefore admissible but not prevalent competitors to the symmetric stationary states of Appendix~\ref{sec:appendix_even_N}.

\bibliography{references_nonstationary_cyclops}

\end{document}